\documentclass[floatfix,aps,floats,pre,10pt]{revtex4-2}
\usepackage{graphics,graphicx,epsfig}
\usepackage{epsf,epstopdf,wrapfig}
\usepackage{array, booktabs, makecell}
\usepackage{threeparttable} 
\usepackage{amsmath,amssymb}
\usepackage{xcolor}
\usepackage{dsfont}

\usepackage{newfloat}
\usepackage{algorithm}
\usepackage[mathscr]{euscript}
\usepackage{algpseudocode}
\usepackage{bm}

\DeclareMathOperator*{\argmin}{\arg\!\min}

\usepackage[utf8]{inputenc}

\usepackage{hyperref}

\begin{document}

\title{Optimizing Reservoir Computing for Reconstructing Ergodic Properties}

\author{Akira Kawano}
\thanks{These two authors contributed equally.}
\affiliation{Biological Physics Theory Unit, OIST Graduate University, Okinawa 904-0495, Japan} 

\author{Ilia Soroka}
\thanks{These two authors contributed equally.}
\affiliation{Department of Physics and Astronomy, VU University Amsterdam, 1081HV Amsterdam, The Netherlands} 

\author{Greg J. Stephens}
\affiliation{Biological Physics Theory Unit, OIST Graduate University, Okinawa 904-0495, Japan}
\affiliation{Department of Physics and Astronomy, VU University Amsterdam, 1081HV Amsterdam, The Netherlands}

\begin{abstract} 
Reservoir computing is a powerful framework for modeling dynamical systems due to its universality and computational efficiency. However, a major challenge is achieving a forecast with accurate long-time statistics, or climate, which is essential for inferring ergodic properties such as Lyapunov exponents. A common approach is to optimize the reservoir’s macroscopic parameters, such as the spectral radius, by maximizing prediction time. However, we show that even predictions accurate over multiple Lyapunov times do not guarantee the correct long-time statistics. Instead, we choose reservoir properties by minimizing the error in the reconstructed invariant distribution (or its projections), which is easily available from data. We demonstrate that this approach reproduces the Lyapunov exponents of model dynamical systems, including the logistic and standard maps, as well as the double pendulum, even with partial observations. We further show that recurrent connections, and resulting reservoir memory, are only required in the partially-observed case. We introduce a temporal scaling which reliably separates system and reservoir dynamics. In the posture time series of the nematode {\textit C. elegans} we show that our approach quantitatively reproduces a chaotic behavioral attractor, but this requires a further constraint on the maximal conditional Lyapunov exponent to ensure the reservoir remains consistently synchronized to the complex biological input.
\end{abstract}

\maketitle  

\section*{Introduction}
Understanding the change of an object or system over time has been essential to our understanding of the world, from Aristotle and the natural place of things, to the planetary motions quantitatively characterized by Newton's universal law of gravity, to modern explorations in complex systems. Living systems in particular, from gene expression to neural dynamics and  animal behavior, exhibit rich nonlinear dynamics that are challenging both to characterize analytically, and to predict.

Despite previous progress resulting in a deep formal understanding of nonlinear dynamical systems (see e.g.~\cite{strogatz2024nonlinear,ott2002chaos}), the challenges of today are qualitatively different. Most significantly, there is an ongoing and dramatic increase in the availability of high-resolution time series measurements. Yet, such data is often from systems for which no equations are known, and/or for which we only have partial measurements of the underlying state space. For example, even the (currently) largest-scale measurement of neural dynamics \cite{manley2024simultaneous} captures only a fraction of the activity in a mammalian brain.

Such novel data is driving new analysis as well as adaptations of previous theory. One example is the use of predictive information to quantify the number of required delays in Takens' embedding \cite{Ahamed2021,Costa2023}. Another is the construction of equations of motion directly from data \cite{crutchfield1987equations,champion2019-SINDy, supekar2023learning}. Indeed, this approach has received substantial recent attention with the additional power of deep networks (see e.g.~\cite{Young&Graham2023}).  But equations of motion are most useful when they are interpretable and this is often challenging in complex systems. In addition, these approaches can generically result in equations which produce unstable trajectories.

Another approach is to estimate important ergodic properties, such as the invariant density or the Lyapunov spectrum, directly from data \cite{packard1980geometry}. This is advantageous as these properties convey useful physical meaning and are generally not obvious from the equations themselves, even when the system is known. Indeed, such an analysis provided direct evidence for a low-dimensional chaotic attractor at the onset of turbulence \cite{brandstater1983low}, and in the posture-scale behavior of the nematode {\em C. elegans} \cite{Ahamed2021}. However, in the typical experimental setting of partial observations, this approach requires estimation of both the embedding dimension and the Jacobian \cite{Wolf1985}, each of which introduces errors that can both produce spurious exponents and miss real ones.
 
Reservoir networks, a class of high-dimensional, deterministic dynamical systems which have their own intrinsic and often chaotic dynamics offer a powerful route for prediction and understanding \cite{platt_systematic_2022, lukovsevivcius2009reservoir, vlachas2020backpropagation}. Typically the dynamics within a reservoir emerge from random but fixed internal connections. The reservoir is then driven with the system of interest. After a sufficient period of driving, predictions for the system are then derived from a linear combination of the reservoir nodes.  In this way, a reservoir's predictive power is conceptually similar to the lottery ticket hypothesis \cite{frankle&carbin}: the innate dynamics of the reservoir are diverse enough to effectively span those of the target system.  While reservoir computing, including specific implementations as Echo State Networks \cite{jaeger2001echo} or Liquid State Machines \cite{maass2002-LiquidStateMachines}, has been in the literature for some time, interest was broadly rekindled when it was shown that reservoirs could produce accurate predictions over many Lyapunov times, as well as the Lyapunov spectrum itself \cite{pathak_using_2017, vlachas2020backpropagation, lu2018attractor}. Reservoirs have been used for prediction across a variety of systems and can even outperform deep neural networks on the prediction of chaotic time series, as, without training internal weights using backpropagation through time, they do not suffer from exploding gradients \cite{mikhaeil2022}.

Theoretically, reservoir computing relies on generalized synchronization, where the reservoir synchronizes with the input\cite{pecora1990synchronization}. Networks that display this synchronization, thus enabling the reservoir to act as an embedding of the driving system, are known to have the Echo State Property \cite{hart2020embedding, grigoryeva2021chaos, hart2024generalised, hart2024attractor, lu2018attractor}. In practice, however, it is not clear how to tune the hyperparameters of the reservoir, which include the spectral radius of matrix of random recurrent connections and the number of reservoir nodes.  Indeed, recent work in learning chaotic dynamics across RNN's has shown that low prediction error does not imply a correct reconstruction of the attractor \cite{hess2023generalized}. Here we address this challenge by optimizing reservoir hyperparameters to minimize the reconstruction error of the invariant distribution, which is directly available from data. We use known dynamical systems with an increasing amount of complexity to show that this approach recovers the correct Lyapunov spectrum, even from partial observations. We also introduce a temporal scaling that can distinguish physical exponents from purely reservoir modes. Finally, we apply our understanding to the posture-scale dynamics of the nematode worm {\em C. elegans} \cite{stephens2008dimensionality}.

\section*{Results}
We first explore reservoir computing in two discrete-time dynamical systems, the logistic map and the standard map.  By using map systems we are able to assess the reservoir's predictive and reconstructive capabilities independently of the discretization required for continuous time dynamics, which can introduce additional complexities. 

\subsection*{Reservoir computing for discrete time maps}
\begin{figure}[ht]
\begin{center} 
\includegraphics[width=0.8\linewidth]{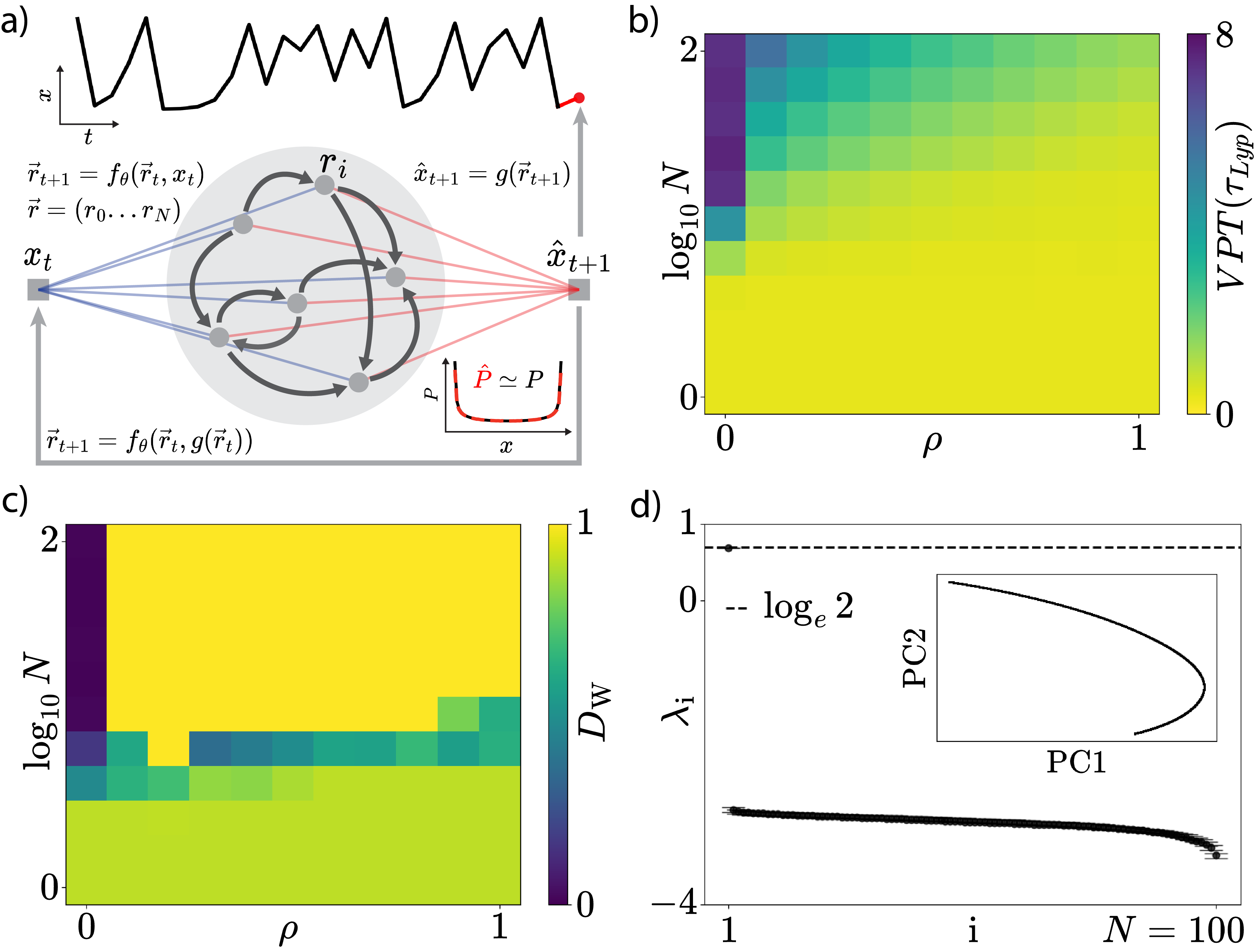}
\caption{{\bf Optimizing reservoir computing via invariant distributions yields the ergodic properties of the logistic map.}
(a) We drive reservoir units $\vec r_t$ via our input data $x_t$ based on a map $f_{\theta}$. Parameters of the reservoir network, such as the spectral radius of the connectivity matrix, are denoted by $\theta$. We use the reservoir response to predict future data, $\hat x_{t+1}=g(\vec r_{t+1})\simeq x_{t+1}$. We then use $\hat x_t$ to construct an autonomous reservoir dynamics, from which we can estimate ergodic properties of the original dynamics, such as the invariant distribution and the Lyapunov spectrum. (b) We measure the time when our prediction error exceeds a fixed threshold, the valid prediction time $VPT$, for networks with a variety of spectral radii $\rho$ and number of neurons $N$. Here, the optimal VPT is for $\rho=0$; recurrent connections are not necessary to predict input with no memory. (c) We use the 1D Wasserstein distance $D_\mathrm{W}$ to quantify the difference between the sample distribution of the input $x_t$ and that of the simulation $\hat x_t$ from networks with various $\rho$ and $N$. Although $D_\mathrm{W}$ is again optimal at $\rho = 0$, the landscape differs from the $VPT$; successful short-time prediction does not guarantee a correct long-term climate. (d) We use optimal parameters to simulate a long trajectory of the autonomous reservoir, and we compute its Lyapunov spectrum. The only positive exponent matches the Lyapunov exponent of the Logistic map with $r=4$. The spectral gap captures the (nonlinear) one dimensional manifold of the input (inset).}
\label{fig:fig1}
\end{center}
\end{figure}

From the general approach of Fig.~\ref{fig:fig1}a we adopt the following specific reservoir architecture and dynamics, 
\begin{equation}
    \vec{r}_{t+1} = \tanh\left( {W}_r \vec{r}_t + {W}_{in} x_t + \vec b \right),
  \label{eq:rc}
\end{equation}
where $\vec{r}_t$ is the instantaneous activity across all $N$ nodes of the network, $x_t$ is an m-dimensional input (typically scaled to zero-mean and unit variance), $W_r$ is a $N \times N$ connectivity matrix, $W_i$ is an $N \times m$ input matrix, and $\vec{b}$ is an N-dimensional bias. We choose the elements of the matrix $W_r$ by randomly sampling from a normal distribution $\mathcal{N}(0, g)$ and we report the strength of the connectivity matrix through the spectral radius $\rho=\max |\lambda_i |$. In the large-N limit this is $\rho=2g\sqrt{N}$. We choose the elements of the matrix $W_{in}$ and the bias $\vec{b}$ by randomly sampling from a normal distribution $\mathcal{N}(0, 1)$. A finite bias is known to have a significant positive impact on forecast performance \cite{platt_systematic_2022}. 

We evolve the reservoir dynamics, Eq.~\ref{eq:rc}, for an initialization time $\tau$. Beyond this time $\tau$, we build a linear prediction from the driven reservoir state to a future input prediction, $\hat x_{t+1}$  
\begin{equation}
\hat x_{t+1}=W_{out}\vec r_{t+1},
\end{equation}
where the $m \times N$ matrix $W_o$ is determined by Tikhonov-regularized (ridge) regression.  We then construct an autonomous reservoir dynamics by using this prediction as input for the next step, i.e., replacing $x_t$ in eq. \ref{eq:rc} with $\hat x_t$, 
\begin{equation}
    \vec{r}_{t+1} = \tanh\left( W_r \vec{r}_t + W_{in} W_{out} \vec{r}_t + \vec b \right)
    \label{eq:autorc}
\end{equation}

\subsection*{Prediction and ergodic properties of the logistic map}
The logistic map, a discrete time nonlinear dynamical system originally devised as simple population model \cite{may1976}, offers an ideal illustration of reservoir computing,
\begin{equation}
x_{n+1}=rx_n(1-x_n)
\label{eq:logistic}
\end{equation}
where $0\leq x \leq1$ is the 1D state space and $0\leq r\leq4$ is a parameter. Beyond $r\sim 3.57$ the map is generally chaotic, though there are isolated regions with regular attractors. In the following we choose $r=4$ which results in a maximally chaotic dynamics. We simulate the logistic map for $T=8\times10^4$ steps and split the trajectory into training and test epochs, each of length $T/2$.

We use the training dynamics to determine $W_o$ and the test dynamics to generate a long trajectory from the autonomous reservoir Eq.~\ref{eq:autorc}. The autonomous dynamics are expected to approximate the long-term behavior (or ``climate'') of the logistic map. Although a reservoir can robustly perform short-term prediction over a wide range of parameters, obtaining a stable climate, which is essential to infer ergodic properties, is challenging \cite{pathak_using_2017, platt_constraining_2023}. This general problem is even more severe in real-world applications with highly nonlinear and noisy time-series, such as in modeling the behavior of biological systems. 

A common approach for obtaining a stable climate is to choose reservoir parameters by optimizing a valid prediction time $VPT$ \cite{platt_systematic_2022}, Eq.~\ref{eq:VPT}. To explore how $VPT$ depends on the network parameters we systematically varied the spectral radius of the connectivity matrix $\rho$ and the number of reservoir neurons $N$, and computed the median $VPT$ from $N_{trial}=30$ realizations of each pair of parameters, Fig \ref{fig:fig1}b. An optimal $VPT\sim8\ \tau_\mathrm{Lyp}$ appeared at $\rho=0$ for $N>16$, illustrating that recurrent connections are not necessary to predict the input. This is because in this example the input time series contains full information, and thus a nonlinear transformation of the current input state without memory is sufficient to capture the dynamics, effectively learning the nonlinearity of the logistic equation, Eq.~\ref{eq:logistic}.

An important challenge in $VPT$-based optimization is the absence of a well-defined prediction scale - that is how large should the prediction time be to guarantee a recovered climate? This absence makes it difficult to robustly assess how close a model is to optimal performance. For instance, we observed that networks with $\rho=0.5$ and $N=100$ showed both relatively large $VPT$ $\sim 3 \ \tau_\mathrm{Lyp}$, and large errors in the estimate Lyapunov exponent, SI Fig.~\ref{fig:mle_err_logistic}. 

We instead optimize the reservoir network by minimizing the reconstruction error of the invariant distribution $P(\hat x)$ obtained from the autonomous dynamics, Eq. \ref{eq:autorc}. In 1D dynamics, as here with the logistic map, we can recover this distribution directly from the data, and in higher dimensional systems we can use whatever projection is available from the measurements.  We evaluate the reconstruction error using a statistical distance between the true $P=P(x)$ and predicted $Q=P(\hat x)$ distributions. For a univariate time series we use the 1D Wasserstein distance $D_\mathrm{W}$
\begin{equation}
D_W(P,Q)=\int |C_P(x)-C_Q(x)|dx 
\label{eq:1D-Wasserstein}
\end{equation}
where $(C_P,C_Q)$ are cumulative distribution functions associated with P and Q respectively.  We measured $D_\mathrm{W}$ between the predicted and true distributions of the logistic map across networks with varying $\rho$ and $N$, Fig \ref{fig:fig1}c. Although small $D_\mathrm{W}$ appeared at parameters that yielded optimal $VPT$, the $D_\mathrm{KS}$ and $VPT$ landscapes are not identical. For example, networks with $\rho = 0.5$ and $N = 100$ exhibited a relatively large $VPT \sim 3\ \tau_\mathrm{Lyp}$, but nevertheless failed to reconstruct the invariant distribution, large $D_\mathrm{W}$. Success in short-term prediction does not guarantee correct long-term forecasts. We note that other distance measures are possible though we don't generally expect qualitative differences. Indeed, in Fig.~S1(a) we show a similar variation in the Kolmogorov-Smirnov distance, the statistic of the Kolmogorov-Smirnov goodness-of-fit test \cite{massey_kolmogorov-smirnov_1951}, across network parameters.  In Fig.~S2 we show the error landscape for the reservoir estimation of the Lyapunov exponent.

We used the optimal reservoir parameters, $\rho=0$ and $N=100$, to simulate a long ($T=10^4$) trajectory of the autonomous dynamics and calculated the Lyapunov spectrum $\lambda_\mathrm i$, Fig \ref{fig:fig1}d (Methods). The only positive exponent matches exactly the Lyapunov exponent of the logistic map, which is $\log_e2$. Moreover, the large spectral gap reflects the dimensionality of the input time series, indicating that only one-dimensional dynamics corresponding to the largest Lyapunov exponent are present in the input. Indeed, the network trajectory projected onto two dominant PCA modes forms a one-dimensional manifold, Fig.~\ref{fig:fig1}d(inset).

\subsection*{Ergodic properties of the standard map from partial observations}

\begin{figure}[h]
\begin{center} 
\includegraphics[width=0.7\linewidth]{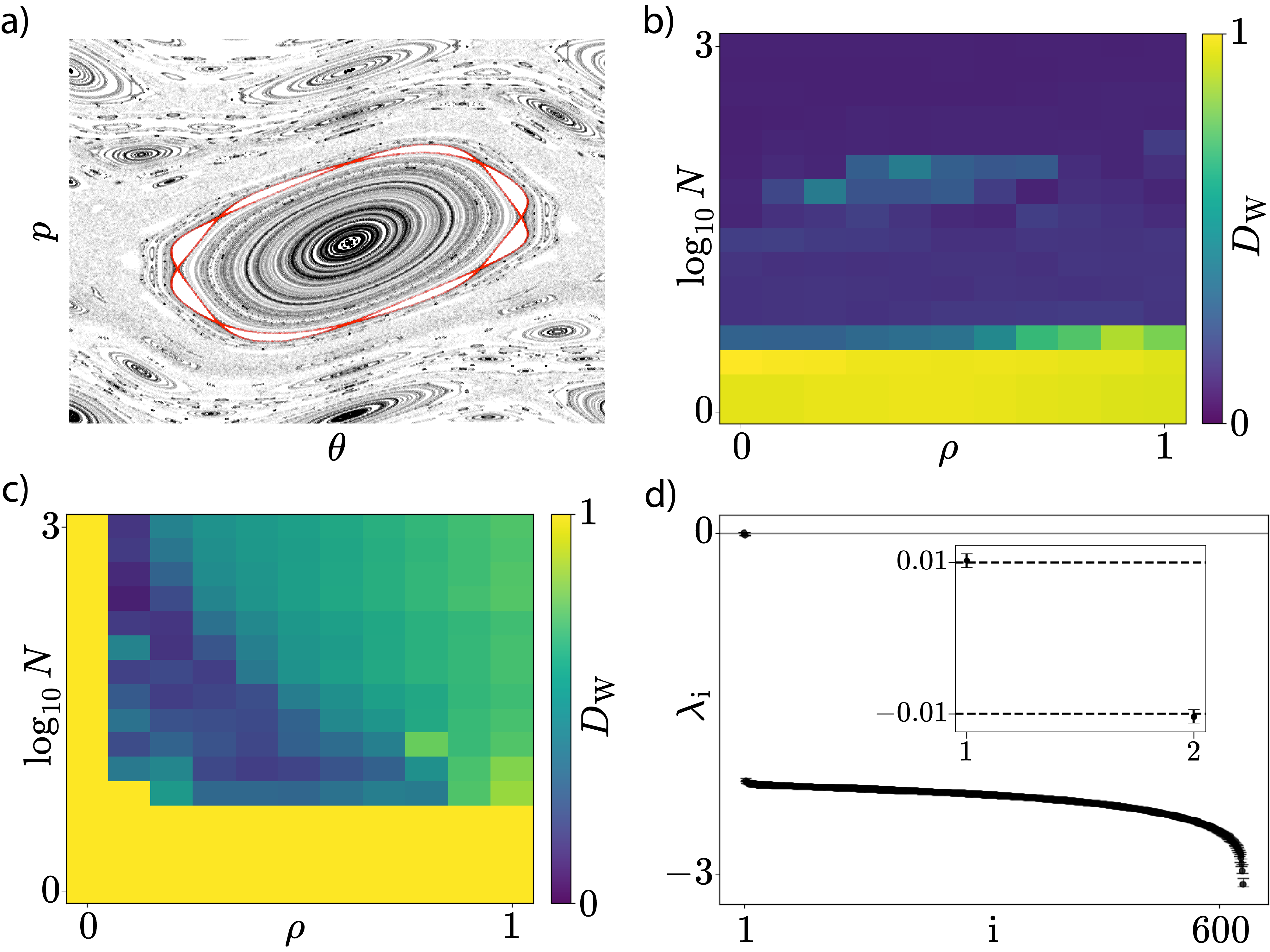}
\caption{{\bf Reproducing ergodic properties of the standard map from a partial observation.} 
(a) The phase space of the standard map with $K=1$ contains fixed points, cycles, and chaotic regions, depending on the initial condition. We adopt an initial condition that will occupy a weakly chaotic region, shown in red.
(b) We first provide full information, $\{\theta, p\}$, to reservoir networks with various values of $\rho$ and $N$. To evaluate distribution reconstruction, we focus on the marginal distribution of $\theta$ and quantify reconstruction accuracy using the Kolmogorov–Smirnov distance $D_\mathrm{KS}$. As in the case of the logistic map, a network with $\rho=0$ achieve optimal performance because memory is unnecessary when the full state is provided.
(c) We next provide only partial observations, $\{\theta\}$, to the networks. In this setting, networks with $\rho=0$ fail to reproduce the distribution, indicating that memory is required to model the underlying input dynamics.
(d) We compute the Lyapunov spectrum of an optimized network trained on partial observations $\{\theta\}$. A spectral gap between indices $\mathrm{i}=2$ and $\mathrm{i}=3$ emerges, indicating that the network recovers the full dimensionality of the standard map. Moreover, the two largest exponents match those of the standard map obtained from its analytical Jacobian, which exhibit the symplectic symmetry $\lambda_1 = -\lambda_2$.}
\label{fig:fig2}
\end{center}
\end{figure}

We next explore reservoir computing in the 2D system of the standard map, originally used to model resonances in Hamiltonian systems \cite{Chirikov1979}, where we can also challenge the reservoir by providing only partial information as input.  Having access to only partial information is a generic challenge when seeking understanding from time series data of complex systems, and was a principal motivation for the development of time-delay embedding \cite{citeulike:2735031}.  Furthermore, the standard map is area-preserving, presenting a discrete analog of time-reversal symmetry in Hamiltonian systems, and thus an additional challenge for reservoir computing \cite{zhang2021learning} climate reconstruction. The standard map dynamics are:
\begin{eqnarray}
\theta_{n+1}&=&\theta_n+p_{n+1}\,\, \mathrm{mod}\ 2\pi  \\ \nonumber
p_{n+1}&=&p_n+K\sin{\theta_n} \,\, \mathrm{mod}\ 2\pi 
\end{eqnarray}
where $K>0$ is a parameter. The phase space for $K=1$ exhibits a mixture of “stable islands” (containing fixed points or cycles) and “chaotic seas” (containing chaotic dynamics) depending on initial conditions, Fig \ref{fig:fig2}a. We adopt an initial condition $(\theta_0, p_0) = (\pi, 4.82)$, which leads to a weakly chaotic dynamics shown in in red.

We first use the complete state information $\{\theta, p\}$ from a trajectory of length $T = 10^5$, dividing it into equal halves to form the training and test sets. The reservoir network (Eq. \ref{eq:rc}) is then driven by the 2D training time series, from which we learn the output matrix $W_o$. Using the resulting autonomous reservoir dynamics (Eq. \ref{eq:autorc}), we estimate the invariant density projected onto the $\theta$ coordinate, $P(\hat\theta)$. Because the marginal density is univariate, as in the previous example, we evaluate its reconstruction relative to the true distribution $P(\theta)$ of the test set using the 1D Wasserstein Distance $D_\mathrm{w}$, Fig. \ref{fig:fig2}b. Networks with $\rho=0$ and $N>100$ achieved optimal performance $D_\mathrm{KS}\sim0$, again suggesting a nonlinear transformation without memory is sufficient to model the complete information input.  As with the logistic map, a similar picture emerges from the Kolmogorov-Smirnov distance, Fig.~S1(b).

We use only $\theta$ variable, i.e.~partial information of the standard map as the input and perform the same training and $D_\mathrm{KS}$ evaluation, Fig. \ref{fig:fig2}c. Under the condition of partial information, networks with $\rho = 0$ can not achieve the invariant-density reconstruction $D_\mathrm{KS} > 0$, indicating that memory is necessary to model the incomplete input time series. Using optimal parameters $\rho=0.1$ and $N=398$ for the partial information input, we calculate the Lyapunov spectrum of the autonomous reservoir dynamics, Fig. \ref{fig:fig2}d. We find a large spectral gap between $\mathrm{i}=2$ and $\mathrm{i}=3$, demonstrating that the network embeds the full dimensionality of the standard map. Furthermore, the two largest exponents match the exponents of the standard map calculated from its analytical Jacobian, including the symplectic symmetry $\lambda_1=-\lambda_2$, Fig.~\ref{fig:fig2}d(inset). Reservoir computing can thus correctly model a dynamical system with symmetries even when only partial observations are available.

\subsection*{Continuous dynamics: the double pendulum}
\begin{figure}[H]
\begin{center} 
\includegraphics[width=0.7\linewidth]{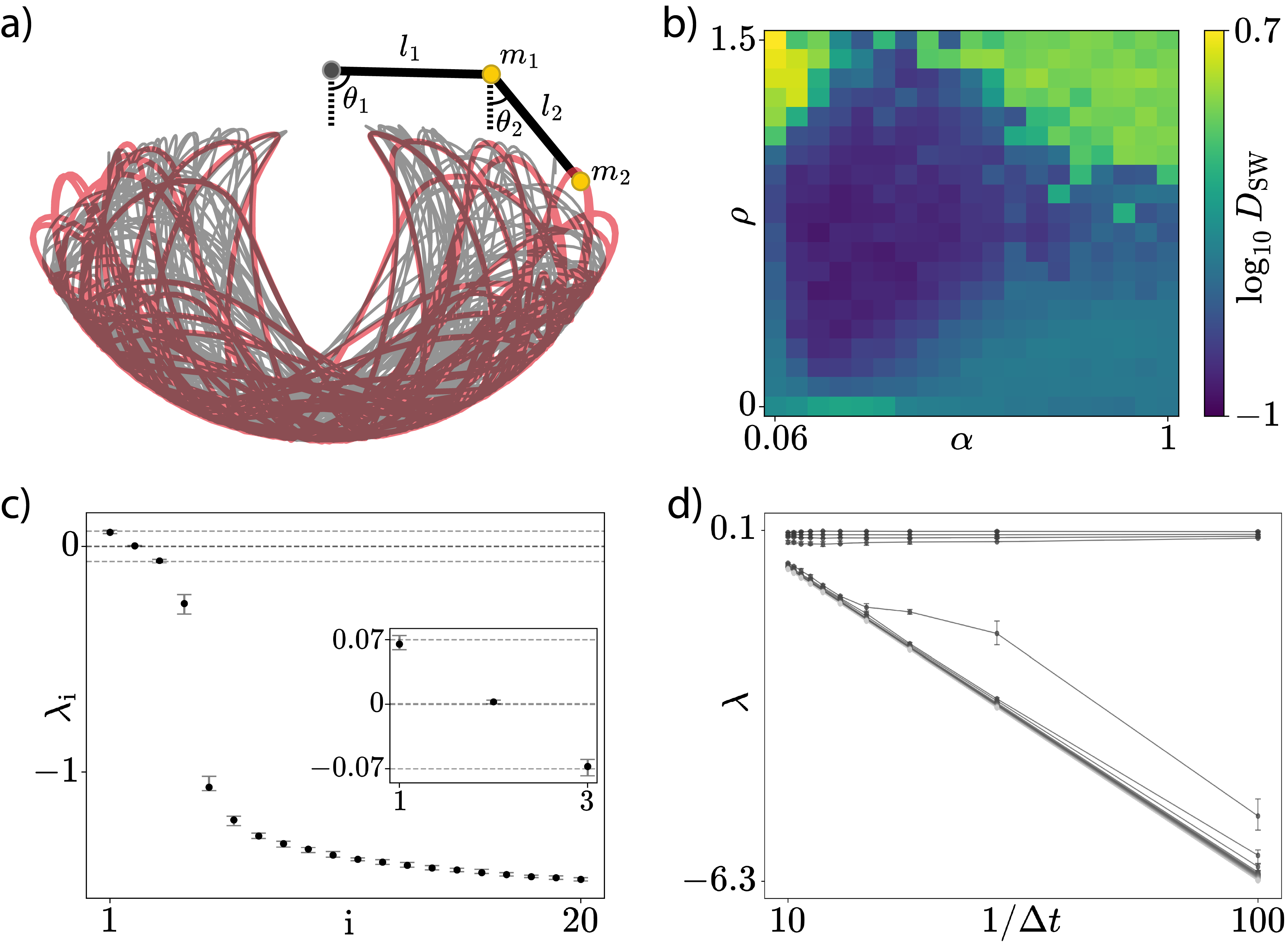}
\caption{{\bf Reservoir dynamics from partial observations of a double pendulum} (a) We show a schematic of the double pendulum as well as the movement of the tip: grey is a numerical simulation while the reservoir prediction is shown in red. (b) Reservoir optimization landscape. We show the median sliced Wasserstein distance, estimated using 50 projections, for the leak rate $\alpha$ and the spectral radius $\rho$ over 30 networks, each tested with 10 different initial conditions. Darker regions indicate lower distance values, representing parameter combinations that yield a more accurate reproduction of the system's dynamics in terms of its invariant distribution.  (c) Lyapunov spectrum of the autonomous reservoir with 700 neurons. We show the first 20 Lyapunov exponents of the reservoir dynamics during autonomous prediction. Black dots with error bars are the exponents calculated from reservoir, while the horizontal dashed gray lines are values calculated using the analytical Jacobian. The error bars represent 95\% confidence intervals, bootstrapped from 10000 distributions sampled with replacement.(d) Within the autonomous reservoir, modes related to the double pendulum are apparent from exponents that are independent of the sampling frequency. In contrast, internal reservoir modes  diverge linearly, $\vec{\lambda}_{internal}\rightarrow-\infty$ (Methods).}
\label{fig:fig3}
\end{center}
\end{figure}

As an example with continuous dynamics, we consider the chaotic double pendulum shown in Fig.~\ref{fig:fig3}(a), where $\theta_1$ and $\theta_2$ are the two angular displacements,  We use equal point masses $m_1=m_2=m$ and equal arm lengths $l_1=l_2=l$, and work in dimensionless time set by $\sqrt{l/g}$ where $g$ is the gravitational acceleration.  The equation of motion are: 
\begin{align}
    \ddot{\theta}_1 &= \frac{-3\sin\theta_1 - \sin(\theta_1 - 2\theta_2) - 2\sin(\theta_1 - \theta_2)\left(\dot{\theta}_2^2 + \cos(\theta_1 - \theta_2)\dot{\theta}_1^2\right)}{3 - \cos(2(\theta_1 - \theta_2))} \\ \nonumber
    \ddot{\theta}_2 &= \frac{2\sin(\theta_1 - \theta_2)\left(2\cos\theta_1 + 2\dot{\theta}_1^2 + \cos(\theta_1 - \theta_2)\dot{\theta}_2^2\right)}{3 - \cos(2(\theta_1 - \theta_2))}
    \label{eq:dp}
\end{align}
As these are two coupled 2nd-order equations, the full phase space is 4-dimensional. The strength of the chaotic dynamics is controlled through the initial conditions and in the following we choose $(\theta_1, \theta_2, \dot{\theta}_1, \dot{\theta}_2) = (0.6, 2.2, 0, 0) \,\,{\rm rad}$. In Fig.~\ref{fig:fig3}(a, black) we show an example endpoint trajectory. 

To better capture the dynamics of this continuous time system, we extend our reservoir architecture to include a leak rate $\alpha \in (0, 1]$ \cite{jaeger2007optimization}, and note that our previous reservoirs are a special case of $\alpha = 1$. The dynamics of the leaky reservoir are
\begin{equation}
    \vec{r}_{t+1} = (1-\alpha)\vec{r}_t+ \alpha\tanh\left( {W}_r \vec{r}_t + {W}_{in}\vec{x}_t + \vec b \right)
  \label{eq:lrc}
\end{equation}
As derived from the Euler discretization of a continuous-time neuron \cite{jaeger2007optimization} the leaking rate introduces an intrinsic time constant to the reservoir that allows us to adjust the timescale to match that of the dynamical system we are trying to learn. Recent work has also shown that such time-history terms enhance the reservoir's delay capacity \cite{ebato2024impact}.

As in the standard map example, we drive the leaky reservoir (Eq.~\ref{eq:lrc}) with partial information, using only the two configuration angles $(\theta_1,\theta_2)$, and learn a linear prediction of the next input state. We then evolve the reservoir autonomously. A typical output trajectory is shown in Fig.~\ref{fig:fig3}(a, red).

We optimize the leaky reservoir architecture over the values of the spectral radius $\rho$ and the leaky rate $\alpha$.  However, to make full use of the available partial information, which will be especially important when we move beyond models, here we use the 2D $\{\theta_1,\theta_2\}$ projection of the full invariant density and the sliced Wasserstein distance \citep{bonneel2015sliced} to compare distributions. The sliced Wasserstein distance is appealing as it approximates optimal transport by averaging the distance between two multidimensional distributions across random one-dimensional projections (Methods).  

In Fig.~\ref{fig:fig3}(b) we show the results of this optimization for a network of $N=700$ reservoir neurons. The input scaling and bias scaling were kept constant at one (consistent with the map examples). The optimal spectral radius was found to be $\rho=0.71$, $\alpha=0.17$ and the resulting median distance $D_{SW}=0.12$.  As with the standard map, the nonzero spectral radius allows reservoir memory for accurate predictions from partial observations.

In Fig.~\ref{fig:fig3}(c) we show the Lyapunov spectrum of the autonomous reservoir dynamics. Despite being trained only on partial state space observations (angles), the reservoir accurately reproduces the largest positive exponent ($\lambda_1$) and the zero exponent ($\lambda_2$) characteristic of the flow. Remarkably, the first three exponents match those computed from the analytical Jacobian, \ref{fig:fig3}(c, inset), including the sympletic symmetry which is visible here as $\lambda_3 \approx -\lambda_1$). The fourth Lyapunov exponent computed from the autonomous reservoir is however, incorrect, and so an important question is whether these dynamics are associated with the system or the reservoir. This is a general concern when the gap in the Lyapunov spectrum is ambiguous.

We introduce a scaling argument that can cleanly separate the system from internal reservoir behavior. The Echo State Property requires the discrete-time reservoir map to be contractive, characterized by a Lipschitz constant $L_F < 1$~\cite{grigoryeva2021chaos}. Consequently, the reservoir possesses its own intrinsic Lyapunov spectrum that imposes a fixed rate of contraction \textit{per iteration}, independent of the physical sampling interval $\Delta t$. Spurious modes arising from this architectural contraction exhibit discrete-time exponents $\mu_\perp \approx \ln(L_F)$; when converted to continuous-time units, $\lambda_\perp = \mu_\perp / \Delta t$ scales as $O(1/\Delta t)$. In contrast, exponents corresponding to the true underlying dynamics satisfy $\mu_i \approx \lambda_i^{\mathrm{true}} \Delta t$~\cite{hart2024generalised, hart2020embedding}, so the estimated physical exponents $\lambda_i = \mu_i/\Delta t \approx \lambda_i^{\mathrm{true}}$ remain invariant as $\Delta t$ varies. Plotting $\lambda_i$ against sampling frequency thus separates physical modes, which appear as horizontal asymptotes, from spurious reservoir modes, which diverge linearly to $-\infty$, Fig.~\ref{fig:fig3}d.

Our scaling argument reveals that the fourth exponent is a genuine feature of the synchronized dynamics, stable under time-step scaling, and distinct from the fast-decaying spurious reservoir modes that appear from index $i=5$ onwards, see also SI Fig.~S3.  However, for a chaotic double pendulum the true Lyapunov spectrum consists of two non-zero exponents which are equal in magnitude but opposite in sign, and two zero exponents, which correspond to the symmetries of continuous dynamics and energy conservation. We note that only the zero associated with continuous dynamics is present in our reservoir spectrum. In fact, the reservoir is exposed to the system dynamics {\em only} on a particular energy hypersurface set by the initial conditions and thus does not directly experience fluctuations off this surface \cite{pathak_using_2017}.

\subsection*{Lyapunov spectrum from data: posture dynamics of the nematode {\em C. elegans}}

Encouraged by the success of our reservoir approach on known systems, including those that are only partially-observed, we explore an example for which there is no analytical model, only time series data: the posture-scale behavior of the nematode worm {\em C.~elegans}. {\em C.~elegans} is an important model system in biology (see e.g.~\cite{zhen2015}), and also especially interesting in the context of dynamical systems as global brain activity can now be measured concurrently with behavior \cite{hallinen2021, atanas2023}.

\begin{figure}[h]
\begin{center} 
\includegraphics[width=0.7\linewidth]{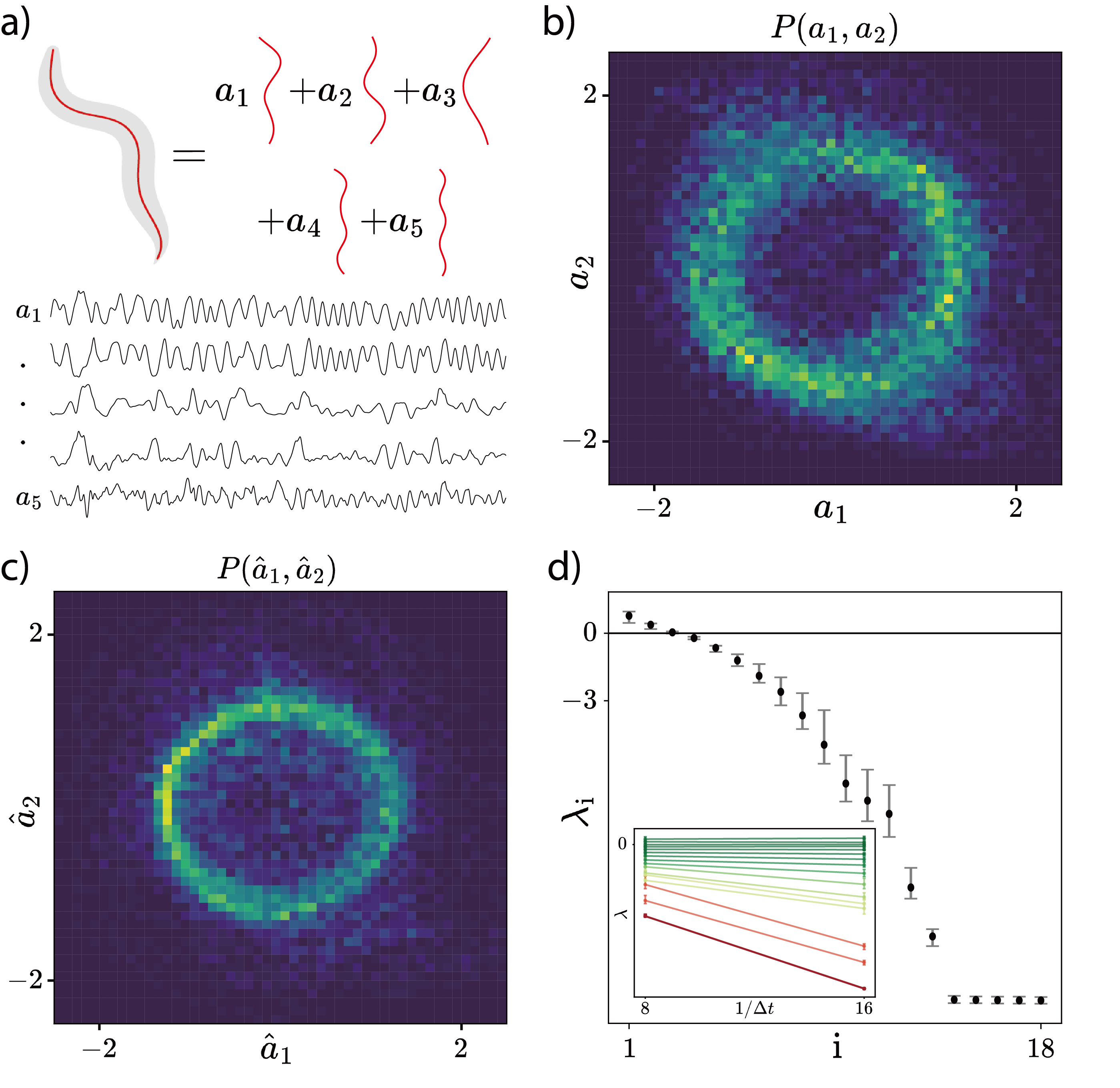}
\caption{{\bf Lyapunov spectrum of {\em C. elegans} posture dynamics from reservoir computing} (a, top) We quantify worm posture as the centerline of high-resolution body images, which we decompose into five eigenworms \cite{stephens2008dimensionality}. (a,bottom) The time series of the eigenworm amplitudes. (b)  We show the joint distribution of the projections onto the leading two eigenworms, $(a_1,a_2)$; these approximately trace the phase of the locomotory crawling wave.  (c) Autonomous reservoir reconstruction of the the leading mode amplitudes, demonstrating that we faithfully reproduce this nontrivial joint distribution. (d) Lyapunov spectrum estimated from the autonomous reservoir dynamics. Error bars are 95\% percentile intervals from a nested bootstrap that propagates both reservoir-selection uncertainty ($m=40$-out-of-$M=100$ without replacement) and trajectory-measurement uncertainty (bootstrap-of-median over 15 independent trajectory realisations of each selected reservoir; $B_{\rm outer}=1000$). (Bottom right) The top 9 exponents are approximately invariant under time-step scaling, suggesting that they reflect physical dynamics, and these are broad agreement with previous work \cite{Ahamed2021}. Exponents 10-12 show intermediate behavior, while higher indices diverge as $O(1/\Delta t)$ and are thus identified with internal reservoir dynamics.  The presence of positive Lyapunov exponents is an indication that chaos is an important component of worm behavior.}
\label{fig:fig4}
\end{center}
\end{figure}

We apply our reservoir approach to the time series of the worm's posture dynamics, Fig.~\ref{fig:fig4}(a). From previously published data (Methods), we pick an exemplar recording and we optimize the reservoir dynamics by minimizing the maximum sliced Wasserstein distance of the distribution of the input posture modes. In this system, however, we consistently found reservoirs with positive conditional Lyapunov exponents (CLE's) which violate the echo state property. CLE's quantify the rates of error divergence conditioned on external forcing and positive values imply that the reservoir under forcing, eq. \ref{eq:rc}, exhibits dynamics unrelated to the input. Such reservoirs are in a state of incomplete synchronization and their Lyapunov exponents do not reflect input dynamics.  While such networks can achieve low distribution reconstruction error, they cannot be used to infer physical Lyapunov exponents \cite{hart2024attractor}. 
Indeed, a requirement for accurate estimation of negative Lyapunov spectrum is that the reservoir's CLE's must be more negative than the most negative exponent of the target system: $\lambda_{\max}^{\text{cond}} < \lambda_{\min}^{\text{target}}$ \cite{hart2024attractor}. This creates a fundamental tension in current reservoir architectures as increasing the spectral radius, which is often necessary for sufficient memory, generally increases the maximum conditional Lyapunov exponent across zero, thus violating the Echo State Property.

For the worm's posture data we thus incorporated the Echo State Property as an explicit constraint in the Bayesian optimization, requiring $\lambda_\mathrm{CLE} <  0$. This constrained optimization successfully identified networks which both match the invariant distribution, for example reproducing the circular distribution of modes $(a_1,a_2)$, Fig.~\ref{fig:fig4}(b,c), and also maintain proper generalized synchronization.  In Fig.~\ref{fig:fig4}(d) we show the Lyapunov spectrum of the autonomous reservoir resulting from the worm's posture dynamics. As with the double pendulum we also show the behavior of these exponents under temporal rescaling, Fig.~\ref{fig:fig4}(d, inset). Physical Lyapunov exponents remain invariant under resampling, while spurious reservoir modes scale as $O(1/\Delta t)$. Exponents 1-9 are invariant and thus intrinsic to the worm's posture dynamics. Exponents 10-12 occupy an intermediate regime: they do not diverge linearly like spurious reservoir modes, but neither are they fully invariant under temporal rescaling. The maximum conditional Lyapunov exponent $\lambda^{\max}_{\mathrm{CLE}}$ sets a resolution floor, exponents more negative than $\lambda^{\max}_{\mathrm{CLE}}$ cannot be reliably distinguished from reservoir contraction. When resampling from 16 to 8~Hz, $\lambda^{\max}_{\mathrm{CLE}}$ shifts from $-16.4$ to $-8.2~\mathrm{s}^{-1}$, approaching the magnitude of exponents 10-12 ($\lambda \approx -8~\mathrm{s}^{-1}$). As the CLE floor rises toward these exponents, they can no longer be resolved as time-step invariant, explaining their intermediate scaling behavior.  In SI Fig.~S5 we show a reconstructed Lyapunov spectrum {\em without} the constraint on the maximal conditional exponent.

Our results are quantitatively consistent with previous findings while revealing additional structure. In particular, Ahamed et al.~\cite{Ahamed2021} reported a seven-dimensional state space dominated by forward, backward, and turning modes, with two positive Lyapunov exponents $\lambda_1 = 0.66 \, (0.62, 0.69) \, {\rm s}^{-1}$ and $\lambda_2 = 0.29\, (0.26, 0.32) \, {\rm s}^{-1}$, and a Kaplan-Yorke dimension $D_{\rm KY} = 5.93 \, (5.75, 6.08)$. Our reservoir-based estimates, $\lambda_1 = 0.79 \, (0.51, 0.96) \, {\rm s}^{-1}$ and $\lambda_2 = 0.38\, (0.20, 0.43)
\, {\rm s}^{-1}$, and $D_{KY}^{\rm reservoir} = 5.25 \pm 0.14$, are in broad agreement for both positive exponents, and we see a similar spectrum across the ensemble of individual recordings, Fig.~S4. Importantly, we find this agreement despite fundamental methodological differences: our approach uses no time-delay embedding or dimensionality reduction. Our reservoir approach may also be more sensitive to dimensions that only subtly influence the dynamics as the scaling analysis reveals $N\sim 12$ modes associated with worm behavior. Resolving whether these additional dimensions reflect physical dynamics requires recordings at higher temporal resolution, which would push the CLE floor more negative and allow a clean separation.  We note that perturbations along these directions decay relatively quickly and are thus less visible in statistical measures. Indeed, unlike classical methods of ergodic estimation from data which require an accurate state-space dimension to avoid spurious Lyapunov exponents~\cite{bryant1990lyapunov}, both dimension and the Lyapunov spectrum are a simultaneous direct output of our reservoir approach.

\section*{Discussion}
In the current era of rapidly advancing progress in time series prediction, the conceptually simple approach of reservoir computing is remarkably powerful, even when compared to state-of-the-art deep recurrent networks, which generally suffer from exploding or vanishing gradients when applied to chaotic systems \cite{mikhaeil2022}.  Here, we used known dynamical systems to probe the use of reservoirs for {\em understanding}, not merely prediction, and we applied this understanding to the data-driven analysis of {\em C. elegans} behavior. Given the time series of a target dynamical system, we showed that it's ergodic properties such as the embedding dimension and Lyapunov spectrum could be inferred from reservoirs constrained to produce accurate projections of the invariant density. Indeed, we find that from the rich internal dynamics of the reservoir, we effectively select a model of the target, and this is especially clear in the prediction of the logistic map, Fig.\ref{fig:fig1}(d). Such reservoirs are also generally good at prediction, though importantly these are not equivalent. Good prediction does not guarantee accurate reconstruction, which is also a generally important caveat for AI interpretability.  

We focus on reservoir computing as this allows direct access to the Jacobian of the dynamical system. We thus transform the model from a statistical estimator into a synthetic dynamical system. Without the Jacobian, we cannot distinguish between a model that has accureately inferred the rules, versus simply parroting the path.

Our approach is similar in spirit to previous work that constrained reservoir networks to match a more complete set of dynamical invariants \cite{platt_constraining_2023}. Here however, we aim for the reservoirs to work directly from data, from which only the invariant density (or it's projections) is relatively easy to estimate. We thus instead use our constrained reservoirs to estimate the Lyapunov spectrum.  

For system identification we note that inferring the Lyapunov spectrum rather than equations of motion can provide more direct insight into the dynamics, unless there is a strong prior on the expected form of the equations. An interesting application for estimating Lyapunov spectra from data lies in the observations of an apparent, general {\em simplicity bias} \cite{dingle2018,dingle2024}. One formal measure of simplicity is provided by Kolmogorov complexity, for which there is an equivalence for dynamical systems of the Kolmogorov-Sinai entropy rate $h_{KS}$, a result known as Brudno's theorem \cite{brudno1982}.  Under general conditions $h_{KS}=\sum \lambda^{+}_i$ \cite{pesin1977characteristic} thus allowing the estimation of the Kolmogorov complexity.  

One of the fundamental challenges of complex systems is the ubiquity of partial observations; we can seldom measure all of the relevant, and coupled degrees of freedom. While in principle this difficulty can be addressed though the use of delay embeddings, determining the number and type of delays can be difficult, though see e.g.~\cite{Costa2023}. For reservoirs, recurrent connections naturally provide memory and thus no delay embedding is required for accurate reconstruction, even with partial observations, as we have shown in the standard map and double pendulum examples 

For determining ergodic properties from data, reservoirs have an important advantage as we need not independently calculate the embedding dimension, for which an incorrect estimation can lead to spurious Lyapunov exponents (see e.g.~\cite{kantz2013}).  For our model systems,  the dimension of the target dynamics is apparent {\em a posteriori} through a substantial gap in the Lyapunov spectrum. Even when no gap is visible, as was the case for the posture time series of {\em C. elegans}, we can determine the dimension through the number of Lyapunov experiments which are invariant under temporal rescaling.  Furthermore, recent theoretical work supports the use of RC for noisy real-world data as the reservoir dynamics have been proven to act as a spectral filter that smoothens input noise \cite{hart2024generalised, duarte2024denoising}, while the high dimensionality of the output layer statistically suppresses error variance via the Central Limit Theorem \cite{hart2024generalised})

While we have focused here on relatively low-dimensional dynamics, we can already draw lessons for more complex systems. First, the continuous, chaotic double pendulum was substantially more difficult than the logistic and standard maps. This is attributable to temporal discretization, which we partially addressed by adding a leak parameter to the reservoir. Even so, we do not see two zero exponents, which we would expect for continuous, energy preserving evolution.  This is likely related to the sampling of the dynamics which is from a single trajectory on an energy hypersurface \cite{pathak_using_2017}. For the posture dynamics of {\em C. elegans}, minimizing the sliced Wasserstein distance between the invariant densities of the reservoir and data alone was insufficient, even for very large reservoirs. Optimization consistently converged to networks violating the Echo State Property, with positive conditional Lyapunov exponents indicating chaotic internal dynamics independent of the input. By incorporating the Echo State Property as an explicit constraint in Bayesian optimization (requiring $\lambda_{max^{cond}} < $ 0), we identified networks that match the invariant distribution and maintain proper generalized synchronization. This constrained optimization approach may prove essential for applying reservoir-based Lyapunov estimation to other noisy, high-dimensional biological systems.  Indeed, our results suggest that for complex, real-world systems, various ``climate'' ergodic properties (such as the the invariant distribution) offer a more robust and physically meaningful fingerprint for model selection than its specific ``weather'' (short-term prediction time). This perspective can be readily applied to other high-dimensional systems where data is limited.

Reservoir computing has been used to model interactions between dynamical systems \cite{huang_detecting_2020}, where the cross-predictability between two reservoirs serves as a measure of coupling. Similar to our study, this approach leverages the parameter-free nature of attractor reconstruction, overcoming a significant hurdle for previous state-space based methods \cite{sugihara_detecting_2012}. Our approach here can also be beneficial for such dual-reservoir settings.

Finally, here we have explored the predictive and modeling capabilities of a fully-connected, random reservoir networks. It will be very interesting for the future to explore modifications to this architecture, such as the addition of gating variables \cite{krishnamurthy2022}.  Understanding the links between computational structure and computation abilities is an important and ongoing direction.

\section*{Acknowledgments}
We thank William Gilpin, Antonio Carlos Costa, Tosif Ahamed and David Jordan as well as members of the Biological Physics Theory Unit for discussions and critical reading. This work was supported by funds from OIST Graduate University (AK, GJS), and Vrije Universiteit Amsterdam (GJS).

\section*{Methods}
\noindent{\bf Software and data availability:} Code for reproducing our results is available here:
\url{https://github.com/oist/ReservoirComputing}. For {\em C. elegans} posture dynamics, the data is available here: \url{https://bitbucket.org/tosifahamed/behavioral-state-space}.  For the analysis of Fig.~\ref{fig:fig4} we used the 2nd timeseries in the file crawl.mat.  We provide an ensemble analysis in SI Fig.~S4.

\medskip 

\noindent {\bf Reservoir Construction:}
The reservoir weight matrix $W_r \in R^{N \times N}$ was constructed as a sparse random matrix with density $d$ and entries drawn from $\mathcal{N}(0, \rho/\sqrt{N})$, where $N$ is the number of units in RC, and zero diagonal. The matrix was then rescaled to achieve a target spectral radius $\rho$:
\begin{equation}
    W_r \leftarrow \frac{\rho}{\rho(W_r)} W_r
\end{equation}
where $\rho(W_r)$ denotes the spectral radius obtained via eigenvalue decomposition (sparse Arnoldi iteration for high-sparsity networks, full decomposition otherwise) using scipy using \texttt{scipy.sparse.linalg.eigs}.

\medskip

\noindent {\bf Reservoir Input Matrix and Bias Construction:} The input matrix $W_{\text{in}} \in R^{N \times D}$, where $D$ is the input dimensionality, has entries sampled from $\mathcal{N}(0, 1)$ for the discrete systems and $\mathcal{U}(-1, 1)$ for continuous, and subsequently rescaled by the input scaling parameter $\sigma_{\text{in}}$. The bias vector $\vec{b} \in R^{N}$ is analogously constructed by sampling from $\mathcal{N}(0, 1)$ or $\mathcal{U}(-1, 1)$ and rescaled by the bias scaling parameter $\sigma_{\text{b}}$.

\medskip 

\noindent {\bf Reservoir Training}: The reservoir state $\vec{r}_t \in R^N$ is initialized from $\mathcal{N}(0, 1)$ and updated according to:
\begin{equation}
    \vec{r}_{t+1} = (\vec{1} - \vec{\alpha}) \odot \vec{r}_t + \vec{\alpha} \odot \tanh\left( W_r \vec{r}_t + W_{\text{in}} \vec{x}_t + \vec{b} \right)
\end{equation}
where $\vec{x}_t \in R^D$ denotes the input at time $t$, $W_r \in R^{N \times N}$ is the reservoir weight matrix, and $\alpha \in (0, 1]$ is the leaking rate. An initial washout period discarded transient responses before collecting reservoir states $R = [\vec{r}_{t_0}, \ldots, \vec{r}_{T}]$. The output weights $W_{\text{out}} \in R^{D \times N}$ and bias $\vec{b}_{\text{out}} \in R^{D}$ are obtained jointly via Tikhonov-regularized regression on the augmented states $\tilde{R} \in R^{(N+1) \times T}$:
\begin{equation}
    [W_{\text{out}}, \vec{b}_{\text{out}}] = Y \tilde{R}^\top \left( \tilde{R} \tilde{R}^\top + \alpha_{\text{reg}} I \right)^{-1}
\end{equation}
where $Y \in \mathbb{R}^{D \times T}$ contains target outputs and $\alpha_{\text{reg}}$ is the regularization coefficient. The linear system was solved efficiently using Cholesky decomposition or scipy SVD solver in case of ill-defined problem.

\medskip 

\noindent {\bf Reservoir Autonomous Prediction:} In autonomous prediction, the reservoir was initialized uniformly in $[-1, 1]$ and driven by a warmup sequence to synchronise with the input dynamics. Following warmup, the network ran in closed-loop:
\begin{equation}
    \vec{\hat{x}}_t = W_{\text{out}} \vec{r}_t + \vec{b}_{\text{out}}
\end{equation}
where $\vec{\hat{x}}_t \in {R}^D$. The prediction was fed back as input to generate the next reservoir state:
\begin{equation}
    \vec{r}_{t+1} = (\vec{1} - \vec{\alpha}) \odot \vec{r}_t + \vec{\alpha} \odot \tanh\left( {W}_r \vec{r}_t + {W}_{\text{in}} \vec{\hat{x}}_t + \vec{b} \right)
\end{equation}
This autonomous iteration was repeated for the desired prediction horizon.

\medskip 

\noindent {\bf Reservoir Lyapunov Spectrum Estimation:} We computed the Lyapunov spectrum of trained ESNs using the standard QR decomposition method \citep{benettin1980lyapunov, eckmann1985ergodic}. The spectrum characterizes the rate of separation of infinitesimally close trajectories in reservoir state space. During autonomous prediction, the closed-loop dynamics become $\vec{r}_{t+1} = f(\vec{r}_t)$, where the output feeds back as input. The Jacobian ${J}_t \in \mathbb{R}^{N \times N}$ describes perturbation evolution:
\begin{equation}
    \vec{\delta}_{t+1} = {J}_t \vec{\delta}_t = (\vec{1} - \vec{\alpha}) \odot \vec{\delta}_t + \vec{\alpha} \odot \left( {D}_t \left( {W}_r + {W}_{\text{in}} {W}_{\text{out}} \right) \vec{\delta}_t \right)
\end{equation}
where $\vec{\delta}_t \in {R}^N$ is the perturbation vector, $\vec{z}_t = {W}_r \vec{r}_t + {W}_{\text{in}} \vec{\hat{x}}_t + \vec{b} \in \mathbb{R}^N$ is the pre-activation, and ${D}_t = \text{diag}(1 - \tanh^2(\vec{z}_t)) \in \mathbb{R}^{N \times N}$. We initialized $k \leq N$ orthonormal perturbation vectors ${\Delta}_0 \in \mathbb{R}^{N \times k}$ and propagated them alongside the reservoir trajectory. Every $\tau$ steps, we performed QR decomposition ${\Delta} = {Q}{R}$, accumulated the logarithms of the diagonal elements of ${R}$, and reorthonormalized. The $i$-th Lyapunov exponent was estimated as:
\begin{equation}
    \lambda_i = \frac{1}{n\tau\Delta t} \sum_{j=1}^{n} \log |R_{ii}^{(j)}|
\end{equation}
where $n$ is the number of reorthonormalizations and $\Delta t$ is the integration timestep.

\medskip 

\noindent {\bf Kolmogorov–Smirnov Distance:} For 1D systems
the Kolmogorov–Smirnov distance $D_\mathrm{KS}$ provides a natural measure to evaluate the accuracy of the reconsructed invariant distribution. $D_\mathrm{KS}$ is the test statistic used in the Kolmogorov–Smirnov goodness-of-fit test \cite{massey_kolmogorov-smirnov_1951} and is defined in the two-sample case as the maximum vertical difference between two cumulative distribution functions, $F(x)$ and $\hat F(x)$
\begin{equation}
    D_\mathrm{KS}=\sup_x\left|F(x)-\hat F(x)\right|
\end{equation}
In addition to its simple definition, the distribution of $D_\mathrm{KS}$ does not depend on either $F(x)$ or $\hat F(x)$. This distribution-free property provides a robust and interpretable measure of reconstruction error. We used the two-sample Kolmogorov–Smirnov statistic implemented in scipy.stats.ks\_2samp from the SciPy library \cite{hodges1958significance}.

\medskip 

\noindent {\bf Sliced Wasserstein Distance} To evaluate reconstruction of the invariant density for systems with $D>1$ we use the Sliced  Wasserstein distance. Recall that for univariate distributions, the 2-Wasserstein distance is:
\begin{equation}
    W_2(P, Q) = \left( \int_0^1 \left| F_P^{-1}(t) - F_Q^{-1}(t) \right|^2 dt \right)^{1/2}
\end{equation}
where $F_P^{-1}$ and $F_Q^{-1}$ are the quantile functions of distributions $P$ and $Q$. Geometrically, this is the root-mean-square of vertical distances between quantile functions, or equivalently, horizontal distances between cumulative distribution functions at matched probability levels. For multivariate data in $\mathbb{R}^D$, we employ the sliced Wasserstein distance \citep{bonneel2015sliced}, which approximates optimal transport by averaging over random one-dimensional projections:
\begin{equation}
    SW_2(P, Q) = \left( \int_{\mathbb{S}^{D-1}} W_2^2(P_\theta, Q_\theta) \, d\sigma(\theta) \right)^{1/2}
\end{equation}
where $\mathbb{S}^{D-1}$ is the unit sphere with uniform measure $\sigma$, and $P_\theta$, $Q_\theta$ denote the distributions of $\theta^\top \mathbf{x}$ for $\mathbf{x} \sim P$ and $\mathbf{x} \sim Q$ respectively.
\medskip

\noindent {\bf Maximum Sliced Wasserstein Distance:} While the sliced Wasserstein distance averages over projection directions, the maximum sliced Wasserstein distance identifies the worst-case projection:
  \begin{equation}
      \mathrm{Max}\text{-}\mathrm{SW}_2(P, Q) = \max_{\theta \in \mathbb{S}^{D-1}} W_2(P_\theta, Q_\theta)
  \end{equation}
  In practice, this is approximated by taking the maximum over a finite set of random projections. By penalizing the single direction with the largest
   distributional mismatch, this metric ensures that no marginal is poorly reconstructed. For the {\em C. elegans} posture dynamics we used the max
  sliced Wasserstein distance with 150 random projections.
\medskip 

\noindent {\bf Valid Prediction Time (VPT):}
We define VPT as the time when the prediction error $e(t)$ exceeds a threshold $\epsilon$ \cite{platt_systematic_2022}
\begin{equation}
    e(t=VPT)>\epsilon
\label{eq:VPT}
\end{equation}
As a measure of prediction error, we use the root mean squared error (RMSE)
\begin{equation}
    e(t)=\sqrt{\frac{1}{d}\sum^d_{i=1}[\hat x_i(t)- x_i(t)]^2}
\end{equation}
We set $\epsilon=0.1$ and evaluate the average of $VPT$ from 100 predictions from different initial conditions for each network realization. 

\medskip 

\noindent {\bf Double Pendulum:} We simulated a double pendulum with equal masses $m_1 = m_2 = 1$ and lengths $l_1 = l_2 = 1$. The equations of motion in terms of angular displacements $\theta_1, \theta_2$ are:

\begin{align*}
    \ddot{\theta}_1 &= \frac{-3\sin\theta_1 - \sin(\theta_1 - 2\theta_2) - 2\sin(\theta_1 - \theta_2)\left(\dot{\theta}_2^2 + \cos(\theta_1 - \theta_2)\dot{\theta}_1^2\right)}{3 - \cos(2(\theta_1 - \theta_2))} \\[6pt]
    \ddot{\theta}_2 &= \frac{2\sin(\theta_1 - \theta_2)\left(2\cos\theta_1 + 2\dot{\theta}_1^2 + \cos(\theta_1 - \theta_2)\dot{\theta}_2^2\right)}{3 - \cos(2(\theta_1 - \theta_2))}
\end{align*}
We evolvled the system using the DOP853 explicit Runge-Kutta method \citep{hairer1993solving} using scipy build in solver, with absolute and relative tolerances of $10^{-13}$. Initial conditions were set to $(\theta_1, \theta_2, \dot{\theta}_1, \dot{\theta}_2) = (0.6, 2.2, 0, 0)$ rad, placing the system in the chaotic regime, simmilar to \cite{zhang2021learning}. Integration was for $T = 5000$ time units with output sampled at $\Delta t = 0.001$, giving $5 \times 10^6$ data points, from which the analytical Lyapunov spectrum was calculated. The resulting dataset was sub-sampled to $\Delta t = 0.1$, giving $5 \times 10^4$ data points for ESN training.

\medskip
\noindent We computed Lyapunov exponents using 
the analytical Jacobian $J(\vec{x}) = \partial\dot{\vec{x}}/\partial \vec{x}$ 
where $\vec{x} = [\theta_1, \theta_2, \omega_1, \omega_2]^T$. Perturbations 
evolved according to the tangent dynamics, integrated via the matrix exponential:
\begin{equation}
    \vec{\delta}_{t+\Delta t} = e^{J(X_t) \Delta t} \vec{\delta}_t
\end{equation}
We initialized $k=4$ orthonormal perturbation vectors and propagated them alongside the trajectory. After a 500-step transient, we performed QR 
decomposition $\Delta = QR$ every $\tau=10$ 
steps, accumulated the logarithms of diagonal elements, and reorthonormalized. 
The $i$-th Lyapunov exponent was estimated as:
\begin{equation}
    \lambda_i = \frac{1}{n\tau\Delta t} \sum_{j=1}^{n} \log |R_{ii}^{(j)}|
\end{equation}
where $n$ is the number of reorthonormalizations and $\Delta t=0.001$ is the 
integration timestep.

\medskip

\noindent {\bf Conditional Lyapunov Exponents:} For the driven system $\vec{r}_{t+1} = g(\vec{r}_t, \vec{x}_t)$, where $\vec{x}_t$ is the driving signal, the conditional Jacobian ${J}_t^{\text{cond}} \in \mathbb{R}^{N \times N}$ describes perturbation evolution without output feedback:
\begin{equation}
    \vec{\delta}_{t+1} = {J}_t^{\text{cond}} \vec{\delta}_t = (\vec{1} - \vec{\alpha}) \odot \vec{\delta}_t + \vec{\alpha} \odot \left( {D}_t {W}_r \vec{\delta}_t \right)
\end{equation}
where $\vec{z}_t = {W}_r \vec{r}_t + {W}_{\text{in}} \vec{x}_t + \vec{b}$ is the pre-activation computed from the input $\vec{x}_t$, and ${D}_t = \text{diag}(1 - \tanh^2(\vec{z}_t))$.

\medskip

\noindent {\bf Double Pendulum Optimization:} For the double pendulum we optimized prediction using grid search over spectral radius $g \in [0.06, 1.5]$ and leaky integration rate $\alpha \in [0.01, 1]$, sampling 19 and 20 values uniformly from each range with 380 total parameter combinations. All ESNs used $N=700$ reservoir neurons with fixed Tikhonov regularization ($\alpha_{\text{reg}}=0.01$), dense connectivity (sparsity=0), and input and bias scaling of one. For each parameter combination, we trained 30 independent ESN realizations with a 1500-step washout period. Each trained network was evaluated on 10 prediction tasks initialized from random points in the training data. After a 1000-step warmup phase, networks generated 4500-step autonomous predictions. 

\medskip 

\noindent {\bf Worm Optimization:} For the prediction of C. elegans posture dynamics, we optimized 
reservoir hyperparameters using single-objective Bayesian optimization with the Ax 
platform. The search space included spectral radius $\rho \in$ [0.01, 3], leaking rate 
$\alpha \in$ [0.01, 1], input scaling $\sigma_{in} \in$ [0.01, 1.5], and bias scaling $\sigma_{b} \in$ [0.01, 1.5]. 
Fixed parameters were N = 10,000 neurons with sparsity 0.99. The generation strategy 
consisted of initial Sobol quasi-random sampling for 20 trials, followed by 
model-based optimization using a Gaussian process surrogate with Expected Improvement 
for 30 trials. We minimized the max sliced Wasserstein distance between predicted and true posture distributions, evaluated with 150 random projections over 8 predictions from 5 independent network realizations. To ensure the Echo State Property, we incorporated $\lambda^{\mathrm{cond}}_{\mathrm{max}} < 0$ 
as an outcome constraint in the Bayesian optimization, where a separate Gaussian 
process surrogate modeled the maximum conditional Lyapunov exponent as a function 
of the hyperparameters, guiding the search toward regions satisfying generalized 
synchronization.

\medskip
\noindent {\bf Ensemble training and uncertainty for the {\em C. elegans} Lyapunov spectrum:} For each of the 12 worm recordings we trained $M=100$ reservoirs at the optimised hyperparameters, differing only in random seed. Each reservoir was scored by the median (over 10 non-overlapping 2000+5000-step windows) of the mean sliced -Wasserstein distance $D_W$ between its autonomous prediction and held-out data, computed with 50 deterministic orthogonal projections shared across all ESNs and worms.

\medskip
\noindent To propagate the uncertainty in selecting the best reservoir, we drew $B=2000$ random subsamples of $m=40$ ESN indices without replacement and identified the $\argmin D_W$ within each. This $m$-out-of-$M$ scheme avoids the known inconsistency of the standard bootstrap for extremum statistics~\cite{bickel2008choice}. For each ESN ever selected as best (13-16 unique per worm), we computed the Lyapunov spectrum from $N_\mathrm{traj}=15$ independent trajectories.

\medskip
\noindent The per-worm confidence interval in Fig.~\ref{fig:fig4}(d) is the 2.5-97.5\% percentile across $B=1000$ iterations, in each of which we drew a 40-index subsample, identified the $\argmin D_W$, resampled its 15 trajectory measurements with replacement, and took the median spectrum. For the across-worm spectrum in Fig.~\ref{fig:S_all}, each of $B=5000$ iterations drew 12 worms with replacement, drew one random spectrum from each selected worm's per-worm distribution, and took the mean across the 12; the reported CI is the percentile interval of these means.
\newpage
\bibliography{references}

\begin{thebibliography}{55}%
\makeatletter
\providecommand \@ifxundefined [1]{%
 \@ifx{#1\undefined}
}%
\providecommand \@ifnum [1]{%
 \ifnum #1\expandafter \@firstoftwo
 \else \expandafter \@secondoftwo
 \fi
}%
\providecommand \@ifx [1]{%
 \ifx #1\expandafter \@firstoftwo
 \else \expandafter \@secondoftwo
 \fi
}%
\providecommand \natexlab [1]{#1}%
\providecommand \enquote  [1]{``#1''}%
\providecommand \bibnamefont  [1]{#1}%
\providecommand \bibfnamefont [1]{#1}%
\providecommand \citenamefont [1]{#1}%
\providecommand \href@noop [0]{\@secondoftwo}%
\providecommand \href [0]{\begingroup \@sanitize@url \@href}%
\providecommand \@href[1]{\@@startlink{#1}\@@href}%
\providecommand \@@href[1]{\endgroup#1\@@endlink}%
\providecommand \@sanitize@url [0]{\catcode `\\12\catcode `\$12\catcode `\&12\catcode `\#12\catcode `\^12\catcode `\_12\catcode `\%12\relax}%
\providecommand \@@startlink[1]{}%
\providecommand \@@endlink[0]{}%
\providecommand \url  [0]{\begingroup\@sanitize@url \@url }%
\providecommand \@url [1]{\endgroup\@href {#1}{\urlprefix }}%
\providecommand \urlprefix  [0]{URL }%
\providecommand \Eprint [0]{\href }%
\providecommand \doibase [0]{https://doi.org/}%
\providecommand \selectlanguage [0]{\@gobble}%
\providecommand \bibinfo  [0]{\@secondoftwo}%
\providecommand \bibfield  [0]{\@secondoftwo}%
\providecommand \translation [1]{[#1]}%
\providecommand \BibitemOpen [0]{}%
\providecommand \bibitemStop [0]{}%
\providecommand \bibitemNoStop [0]{.\EOS\space}%
\providecommand \EOS [0]{\spacefactor3000\relax}%
\providecommand \BibitemShut  [1]{\csname bibitem#1\endcsname}%
\let\auto@bib@innerbib\@empty
\bibitem [{\citenamefont {Strogatz}(2024)}]{strogatz2024nonlinear}%
  \BibitemOpen
  \bibfield  {author} {\bibinfo {author} {\bibfnamefont {S.~H.}\ \bibnamefont {Strogatz}},\ }\href@noop {} {\emph {\bibinfo {title} {Nonlinear dynamics and chaos: with applications to physics, biology, chemistry, and engineering}}}\ (\bibinfo  {publisher} {Chapman and Hall/CRC},\ \bibinfo {year} {2024})\BibitemShut {NoStop}%
\bibitem [{\citenamefont {Ott}(2002)}]{ott2002chaos}%
  \BibitemOpen
  \bibfield  {author} {\bibinfo {author} {\bibfnamefont {E.}~\bibnamefont {Ott}},\ }\href@noop {} {\emph {\bibinfo {title} {Chaos in dynamical systems}}}\ (\bibinfo  {publisher} {Cambridge university press},\ \bibinfo {year} {2002})\BibitemShut {NoStop}%
\bibitem [{\citenamefont {Manley}\ \emph {et~al.}(2024)\citenamefont {Manley}, \citenamefont {Lu}, \citenamefont {Barber}, \citenamefont {Demas}, \citenamefont {Kim}, \citenamefont {Meyer}, \citenamefont {Traub},\ and\ \citenamefont {Vaziri}}]{manley2024simultaneous}%
  \BibitemOpen
  \bibfield  {author} {\bibinfo {author} {\bibfnamefont {J.}~\bibnamefont {Manley}}, \bibinfo {author} {\bibfnamefont {S.}~\bibnamefont {Lu}}, \bibinfo {author} {\bibfnamefont {K.}~\bibnamefont {Barber}}, \bibinfo {author} {\bibfnamefont {J.}~\bibnamefont {Demas}}, \bibinfo {author} {\bibfnamefont {H.}~\bibnamefont {Kim}}, \bibinfo {author} {\bibfnamefont {D.}~\bibnamefont {Meyer}}, \bibinfo {author} {\bibfnamefont {F.~M.}\ \bibnamefont {Traub}},\ and\ \bibinfo {author} {\bibfnamefont {A.}~\bibnamefont {Vaziri}},\ }\bibfield  {title} {\bibinfo {title} {Simultaneous, cortex-wide dynamics of up to 1 million neurons reveal unbounded scaling of dimensionality with neuron number},\ }\href@noop {} {\bibfield  {journal} {\bibinfo  {journal} {Neuron}\ }\textbf {\bibinfo {volume} {112}},\ \bibinfo {pages} {1694} (\bibinfo {year} {2024})}\BibitemShut {NoStop}%
\bibitem [{\citenamefont {Ahamed}\ \emph {et~al.}(2021)\citenamefont {Ahamed}, \citenamefont {Costa},\ and\ \citenamefont {Stephens}}]{Ahamed2021}%
  \BibitemOpen
  \bibfield  {author} {\bibinfo {author} {\bibfnamefont {T.}~\bibnamefont {Ahamed}}, \bibinfo {author} {\bibfnamefont {A.~C.}\ \bibnamefont {Costa}},\ and\ \bibinfo {author} {\bibfnamefont {G.~J.}\ \bibnamefont {Stephens}},\ }\bibfield  {title} {\bibinfo {title} {{Capturing the continuous complexity of behaviour in Caenorhabditis elegans}},\ }\href {https://doi.org/10.1038/s41567-020-01036-8} {\bibfield  {journal} {\bibinfo  {journal} {Nature Physics}\ }\textbf {\bibinfo {volume} {17}},\ \bibinfo {pages} {275} (\bibinfo {year} {2021})}\BibitemShut {NoStop}%
\bibitem [{\citenamefont {Costa}\ \emph {et~al.}(2023)\citenamefont {Costa}, \citenamefont {Ahamed}, \citenamefont {Jordan},\ and\ \citenamefont {Stephens}}]{Costa2023}%
  \BibitemOpen
  \bibfield  {author} {\bibinfo {author} {\bibfnamefont {A.~C.}\ \bibnamefont {Costa}}, \bibinfo {author} {\bibfnamefont {T.}~\bibnamefont {Ahamed}}, \bibinfo {author} {\bibfnamefont {D.}~\bibnamefont {Jordan}},\ and\ \bibinfo {author} {\bibfnamefont {G.~J.}\ \bibnamefont {Stephens}},\ }\bibfield  {title} {\bibinfo {title} {{Maximally predictive states: From partial observations to long timescales}},\ }\href {https://doi.org/10.1063/5.0129398} {\bibfield  {journal} {\bibinfo  {journal} {Chaos: An Interdisciplinary Journal of Nonlinear Science}\ }\textbf {\bibinfo {volume} {33}},\ \bibinfo {pages} {023136} (\bibinfo {year} {2023})}\BibitemShut {NoStop}%
\bibitem [{\citenamefont {Crutchfield}\ and\ \citenamefont {McNamara}(1987)}]{crutchfield1987equations}%
  \BibitemOpen
  \bibfield  {author} {\bibinfo {author} {\bibfnamefont {J.~P.}\ \bibnamefont {Crutchfield}}\ and\ \bibinfo {author} {\bibfnamefont {B.}~\bibnamefont {McNamara}},\ }\bibfield  {title} {\bibinfo {title} {Equations of motion from a data series ‘},\ }\href@noop {} {\bibfield  {journal} {\bibinfo  {journal} {Complex systems}\ }\textbf {\bibinfo {volume} {1}},\ \bibinfo {pages} {417} (\bibinfo {year} {1987})}\BibitemShut {NoStop}%
\bibitem [{\citenamefont {Champion}\ \emph {et~al.}(2019)\citenamefont {Champion}, \citenamefont {Lusch}, \citenamefont {Kutz},\ and\ \citenamefont {Brunton}}]{champion2019-SINDy}%
  \BibitemOpen
  \bibfield  {author} {\bibinfo {author} {\bibfnamefont {K.}~\bibnamefont {Champion}}, \bibinfo {author} {\bibfnamefont {B.}~\bibnamefont {Lusch}}, \bibinfo {author} {\bibfnamefont {J.~N.}\ \bibnamefont {Kutz}},\ and\ \bibinfo {author} {\bibfnamefont {S.~L.}\ \bibnamefont {Brunton}},\ }\bibfield  {title} {\bibinfo {title} {Data-driven discovery of coordinates and governing equations},\ }\href@noop {} {\bibfield  {journal} {\bibinfo  {journal} {Proceedings of the National Academy of Sciences}\ }\textbf {\bibinfo {volume} {116}},\ \bibinfo {pages} {22445} (\bibinfo {year} {2019})}\BibitemShut {NoStop}%
\bibitem [{\citenamefont {Supekar}\ \emph {et~al.}(2023)\citenamefont {Supekar}, \citenamefont {Song}, \citenamefont {Hastewell}, \citenamefont {Choi}, \citenamefont {Mietke},\ and\ \citenamefont {Dunkel}}]{supekar2023learning}%
  \BibitemOpen
  \bibfield  {author} {\bibinfo {author} {\bibfnamefont {R.}~\bibnamefont {Supekar}}, \bibinfo {author} {\bibfnamefont {B.}~\bibnamefont {Song}}, \bibinfo {author} {\bibfnamefont {A.}~\bibnamefont {Hastewell}}, \bibinfo {author} {\bibfnamefont {G.~P.}\ \bibnamefont {Choi}}, \bibinfo {author} {\bibfnamefont {A.}~\bibnamefont {Mietke}},\ and\ \bibinfo {author} {\bibfnamefont {J.}~\bibnamefont {Dunkel}},\ }\bibfield  {title} {\bibinfo {title} {Learning hydrodynamic equations for active matter from particle simulations and experiments},\ }\href@noop {} {\bibfield  {journal} {\bibinfo  {journal} {Proceedings of the National Academy of Sciences}\ }\textbf {\bibinfo {volume} {120}},\ \bibinfo {pages} {e2206994120} (\bibinfo {year} {2023})}\BibitemShut {NoStop}%
\bibitem [{\citenamefont {Young}\ and\ \citenamefont {Graham}(2023)}]{Young&Graham2023}%
  \BibitemOpen
  \bibfield  {author} {\bibinfo {author} {\bibfnamefont {C.~D.}\ \bibnamefont {Young}}\ and\ \bibinfo {author} {\bibfnamefont {M.~D.}\ \bibnamefont {Graham}},\ }\bibfield  {title} {\bibinfo {title} {Deep learning delay coordinate dynamics for chaotic attractors from partial observable data},\ }\href {https://doi.org/10.1103/PhysRevE.107.034215} {\bibfield  {journal} {\bibinfo  {journal} {Phys. Rev. E}\ }\textbf {\bibinfo {volume} {107}},\ \bibinfo {pages} {034215} (\bibinfo {year} {2023})}\BibitemShut {NoStop}%
\bibitem [{\citenamefont {Packard}\ \emph {et~al.}(1980)\citenamefont {Packard}, \citenamefont {Crutchfield}, \citenamefont {Farmer},\ and\ \citenamefont {Shaw}}]{packard1980geometry}%
  \BibitemOpen
  \bibfield  {author} {\bibinfo {author} {\bibfnamefont {N.~H.}\ \bibnamefont {Packard}}, \bibinfo {author} {\bibfnamefont {J.~P.}\ \bibnamefont {Crutchfield}}, \bibinfo {author} {\bibfnamefont {J.~D.}\ \bibnamefont {Farmer}},\ and\ \bibinfo {author} {\bibfnamefont {R.~S.}\ \bibnamefont {Shaw}},\ }\bibfield  {title} {\bibinfo {title} {Geometry from a time series},\ }\href@noop {} {\bibfield  {journal} {\bibinfo  {journal} {Physical review letters}\ }\textbf {\bibinfo {volume} {45}},\ \bibinfo {pages} {712} (\bibinfo {year} {1980})}\BibitemShut {NoStop}%
\bibitem [{\citenamefont {Brandst{\"a}ter}\ \emph {et~al.}(1983)\citenamefont {Brandst{\"a}ter}, \citenamefont {Swift}, \citenamefont {Swinney}, \citenamefont {Wolf}, \citenamefont {Farmer}, \citenamefont {Jen},\ and\ \citenamefont {Crutchfield}}]{brandstater1983low}%
  \BibitemOpen
  \bibfield  {author} {\bibinfo {author} {\bibfnamefont {A.}~\bibnamefont {Brandst{\"a}ter}}, \bibinfo {author} {\bibfnamefont {J.}~\bibnamefont {Swift}}, \bibinfo {author} {\bibfnamefont {H.~L.}\ \bibnamefont {Swinney}}, \bibinfo {author} {\bibfnamefont {A.}~\bibnamefont {Wolf}}, \bibinfo {author} {\bibfnamefont {J.~D.}\ \bibnamefont {Farmer}}, \bibinfo {author} {\bibfnamefont {E.}~\bibnamefont {Jen}},\ and\ \bibinfo {author} {\bibfnamefont {P.}~\bibnamefont {Crutchfield}},\ }\bibfield  {title} {\bibinfo {title} {Low-dimensional chaos in a hydrodynamic system},\ }\href@noop {} {\bibfield  {journal} {\bibinfo  {journal} {Physical Review Letters}\ }\textbf {\bibinfo {volume} {51}},\ \bibinfo {pages} {1442} (\bibinfo {year} {1983})}\BibitemShut {NoStop}%
\bibitem [{\citenamefont {Wolf}\ \emph {et~al.}(1985)\citenamefont {Wolf}, \citenamefont {Swift}, \citenamefont {Swinney},\ and\ \citenamefont {Vastano}}]{Wolf1985}%
  \BibitemOpen
  \bibfield  {author} {\bibinfo {author} {\bibfnamefont {A.}~\bibnamefont {Wolf}}, \bibinfo {author} {\bibfnamefont {J.~B.}\ \bibnamefont {Swift}}, \bibinfo {author} {\bibfnamefont {H.~L.}\ \bibnamefont {Swinney}},\ and\ \bibinfo {author} {\bibfnamefont {J.~A.}\ \bibnamefont {Vastano}},\ }\bibfield  {title} {\bibinfo {title} {Determining lyapunov exponents from a time series},\ }\href {https://doi.org/10.1016/0167-2789(85)90011-9} {\bibfield  {journal} {\bibinfo  {journal} {Physica D: Nonlinear Phenomena}\ }\textbf {\bibinfo {volume} {16}},\ \bibinfo {pages} {285} (\bibinfo {year} {1985})}\BibitemShut {NoStop}%
\bibitem [{\citenamefont {Platt}\ \emph {et~al.}(2022)\citenamefont {Platt}, \citenamefont {Penny}, \citenamefont {Smith}, \citenamefont {Chen},\ and\ \citenamefont {Abarbanel}}]{platt_systematic_2022}%
  \BibitemOpen
  \bibfield  {author} {\bibinfo {author} {\bibfnamefont {J.~A.}\ \bibnamefont {Platt}}, \bibinfo {author} {\bibfnamefont {S.~G.}\ \bibnamefont {Penny}}, \bibinfo {author} {\bibfnamefont {T.~A.}\ \bibnamefont {Smith}}, \bibinfo {author} {\bibfnamefont {T.-C.}\ \bibnamefont {Chen}},\ and\ \bibinfo {author} {\bibfnamefont {H.~D.}\ \bibnamefont {Abarbanel}},\ }\bibfield  {title} {\bibinfo {title} {A systematic exploration of reservoir computing for forecasting complex spatiotemporal dynamics},\ }\href {https://doi.org/10.1016/j.neunet.2022.06.025} {\bibfield  {journal} {\bibinfo  {journal} {Neural Networks}\ }\textbf {\bibinfo {volume} {153}},\ \bibinfo {pages} {530} (\bibinfo {year} {2022})}\BibitemShut {NoStop}%
\bibitem [{\citenamefont {Luko{\v{s}}evi{\v{c}}ius}\ and\ \citenamefont {Jaeger}(2009)}]{lukovsevivcius2009reservoir}%
  \BibitemOpen
  \bibfield  {author} {\bibinfo {author} {\bibfnamefont {M.}~\bibnamefont {Luko{\v{s}}evi{\v{c}}ius}}\ and\ \bibinfo {author} {\bibfnamefont {H.}~\bibnamefont {Jaeger}},\ }\bibfield  {title} {\bibinfo {title} {Reservoir computing approaches to recurrent neural network training},\ }\href@noop {} {\bibfield  {journal} {\bibinfo  {journal} {Computer science review}\ }\textbf {\bibinfo {volume} {3}},\ \bibinfo {pages} {127} (\bibinfo {year} {2009})}\BibitemShut {NoStop}%
\bibitem [{\citenamefont {Vlachas}\ \emph {et~al.}(2020)\citenamefont {Vlachas}, \citenamefont {Pathak}, \citenamefont {Hunt}, \citenamefont {Sapsis}, \citenamefont {Girvan}, \citenamefont {Ott},\ and\ \citenamefont {Koumoutsakos}}]{vlachas2020backpropagation}%
  \BibitemOpen
  \bibfield  {author} {\bibinfo {author} {\bibfnamefont {P.-R.}\ \bibnamefont {Vlachas}}, \bibinfo {author} {\bibfnamefont {J.}~\bibnamefont {Pathak}}, \bibinfo {author} {\bibfnamefont {B.~R.}\ \bibnamefont {Hunt}}, \bibinfo {author} {\bibfnamefont {T.~P.}\ \bibnamefont {Sapsis}}, \bibinfo {author} {\bibfnamefont {M.}~\bibnamefont {Girvan}}, \bibinfo {author} {\bibfnamefont {E.}~\bibnamefont {Ott}},\ and\ \bibinfo {author} {\bibfnamefont {P.}~\bibnamefont {Koumoutsakos}},\ }\bibfield  {title} {\bibinfo {title} {Backpropagation algorithms and reservoir computing in recurrent neural networks for the forecasting of complex spatiotemporal dynamics},\ }\href@noop {} {\bibfield  {journal} {\bibinfo  {journal} {Neural Networks}\ }\textbf {\bibinfo {volume} {126}},\ \bibinfo {pages} {191} (\bibinfo {year} {2020})}\BibitemShut {NoStop}%
\bibitem [{\citenamefont {Frankle}\ and\ \citenamefont {Carbin}(2019)}]{frankle&carbin}%
  \BibitemOpen
  \bibfield  {author} {\bibinfo {author} {\bibfnamefont {J.}~\bibnamefont {Frankle}}\ and\ \bibinfo {author} {\bibfnamefont {M.}~\bibnamefont {Carbin}},\ }\bibfield  {title} {\bibinfo {title} {The lottery ticket hypothesis: Finding sparse, trainable neural networks},\ }in\ \href@noop {} {\emph {\bibinfo {booktitle} {7th International Conference on Learning Representations}}}\ (\bibinfo {year} {2019})\BibitemShut {NoStop}%
\bibitem [{\citenamefont {Jaeger}(2001)}]{jaeger2001echo}%
  \BibitemOpen
  \bibfield  {author} {\bibinfo {author} {\bibfnamefont {H.}~\bibnamefont {Jaeger}},\ }\bibfield  {title} {\bibinfo {title} {The “echo state” approach to analysing and training recurrent neural networks-with an erratum note},\ }\href@noop {} {\bibfield  {journal} {\bibinfo  {journal} {Bonn, Germany: German national research center for information technology gmd technical report}\ }\textbf {\bibinfo {volume} {148}},\ \bibinfo {pages} {13} (\bibinfo {year} {2001})}\BibitemShut {NoStop}%
\bibitem [{\citenamefont {Maass}\ \emph {et~al.}(2002)\citenamefont {Maass}, \citenamefont {Natschl{\"a}ger},\ and\ \citenamefont {Markram}}]{maass2002-LiquidStateMachines}%
  \BibitemOpen
  \bibfield  {author} {\bibinfo {author} {\bibfnamefont {W.}~\bibnamefont {Maass}}, \bibinfo {author} {\bibfnamefont {T.}~\bibnamefont {Natschl{\"a}ger}},\ and\ \bibinfo {author} {\bibfnamefont {H.}~\bibnamefont {Markram}},\ }\bibfield  {title} {\bibinfo {title} {Real-time computing without stable states: A new framework for neural computation based on perturbations},\ }\href@noop {} {\bibfield  {journal} {\bibinfo  {journal} {Neural computation}\ }\textbf {\bibinfo {volume} {14}},\ \bibinfo {pages} {2531} (\bibinfo {year} {2002})}\BibitemShut {NoStop}%
\bibitem [{\citenamefont {Pathak}\ \emph {et~al.}(2017)\citenamefont {Pathak}, \citenamefont {Lu}, \citenamefont {Hunt}, \citenamefont {Girvan},\ and\ \citenamefont {Ott}}]{pathak_using_2017}%
  \BibitemOpen
  \bibfield  {author} {\bibinfo {author} {\bibfnamefont {J.}~\bibnamefont {Pathak}}, \bibinfo {author} {\bibfnamefont {Z.}~\bibnamefont {Lu}}, \bibinfo {author} {\bibfnamefont {B.~R.}\ \bibnamefont {Hunt}}, \bibinfo {author} {\bibfnamefont {M.}~\bibnamefont {Girvan}},\ and\ \bibinfo {author} {\bibfnamefont {E.}~\bibnamefont {Ott}},\ }\bibfield  {title} {\bibinfo {title} {Using machine learning to replicate chaotic attractors and calculate {Lyapunov} exponents from data},\ }\href {https://doi.org/10.1063/1.5010300} {\bibfield  {journal} {\bibinfo  {journal} {Chaos: An Interdisciplinary Journal of Nonlinear Science}\ }\textbf {\bibinfo {volume} {27}},\ \bibinfo {pages} {121102} (\bibinfo {year} {2017})}\BibitemShut {NoStop}%
\bibitem [{\citenamefont {Lu}\ \emph {et~al.}(2018)\citenamefont {Lu}, \citenamefont {Hunt},\ and\ \citenamefont {Ott}}]{lu2018attractor}%
  \BibitemOpen
  \bibfield  {author} {\bibinfo {author} {\bibfnamefont {Z.}~\bibnamefont {Lu}}, \bibinfo {author} {\bibfnamefont {B.~R.}\ \bibnamefont {Hunt}},\ and\ \bibinfo {author} {\bibfnamefont {E.}~\bibnamefont {Ott}},\ }\bibfield  {title} {\bibinfo {title} {Attractor reconstruction by machine learning},\ }\href@noop {} {\bibfield  {journal} {\bibinfo  {journal} {Chaos: An Interdisciplinary Journal of Nonlinear Science}\ }\textbf {\bibinfo {volume} {28}} (\bibinfo {year} {2018})}\BibitemShut {NoStop}%
\bibitem [{\citenamefont {Mikhaeil}\ \emph {et~al.}(2022)\citenamefont {Mikhaeil}, \citenamefont {Monfared},\ and\ \citenamefont {Durstewitz}}]{mikhaeil2022}%
  \BibitemOpen
  \bibfield  {author} {\bibinfo {author} {\bibfnamefont {J.}~\bibnamefont {Mikhaeil}}, \bibinfo {author} {\bibfnamefont {Z.}~\bibnamefont {Monfared}},\ and\ \bibinfo {author} {\bibfnamefont {D.}~\bibnamefont {Durstewitz}},\ }\bibfield  {title} {\bibinfo {title} {On the difficulty of learning chaotic dynamics with rnns},\ }\href@noop {} {\bibfield  {journal} {\bibinfo  {journal} {Advances in neural information processing systems}\ }\textbf {\bibinfo {volume} {35}},\ \bibinfo {pages} {11297} (\bibinfo {year} {2022})}\BibitemShut {NoStop}%
\bibitem [{\citenamefont {Pecora}\ and\ \citenamefont {Carroll}(1990)}]{pecora1990synchronization}%
  \BibitemOpen
  \bibfield  {author} {\bibinfo {author} {\bibfnamefont {L.~M.}\ \bibnamefont {Pecora}}\ and\ \bibinfo {author} {\bibfnamefont {T.~L.}\ \bibnamefont {Carroll}},\ }\bibfield  {title} {\bibinfo {title} {Synchronization in chaotic systems},\ }\href@noop {} {\bibfield  {journal} {\bibinfo  {journal} {Physical review letters}\ }\textbf {\bibinfo {volume} {64}},\ \bibinfo {pages} {821} (\bibinfo {year} {1990})}\BibitemShut {NoStop}%
\bibitem [{\citenamefont {Hart}\ \emph {et~al.}(2020)\citenamefont {Hart}, \citenamefont {Hook},\ and\ \citenamefont {Dawes}}]{hart2020embedding}%
  \BibitemOpen
  \bibfield  {author} {\bibinfo {author} {\bibfnamefont {A.}~\bibnamefont {Hart}}, \bibinfo {author} {\bibfnamefont {J.}~\bibnamefont {Hook}},\ and\ \bibinfo {author} {\bibfnamefont {J.}~\bibnamefont {Dawes}},\ }\bibfield  {title} {\bibinfo {title} {Embedding and approximation theorems for echo state networks},\ }\href@noop {} {\bibfield  {journal} {\bibinfo  {journal} {Neural Networks}\ }\textbf {\bibinfo {volume} {128}},\ \bibinfo {pages} {234} (\bibinfo {year} {2020})}\BibitemShut {NoStop}%
\bibitem [{\citenamefont {Grigoryeva}\ \emph {et~al.}(2021)\citenamefont {Grigoryeva}, \citenamefont {Hart},\ and\ \citenamefont {Ortega}}]{grigoryeva2021chaos}%
  \BibitemOpen
  \bibfield  {author} {\bibinfo {author} {\bibfnamefont {L.}~\bibnamefont {Grigoryeva}}, \bibinfo {author} {\bibfnamefont {A.}~\bibnamefont {Hart}},\ and\ \bibinfo {author} {\bibfnamefont {J.-P.}\ \bibnamefont {Ortega}},\ }\bibfield  {title} {\bibinfo {title} {Chaos on compact manifolds: Differentiable synchronizations beyond the takens theorem},\ }\href@noop {} {\bibfield  {journal} {\bibinfo  {journal} {Physical Review E}\ }\textbf {\bibinfo {volume} {103}},\ \bibinfo {pages} {062204} (\bibinfo {year} {2021})}\BibitemShut {NoStop}%
\bibitem [{\citenamefont {Hart}(2024{\natexlab{a}})}]{hart2024generalised}%
  \BibitemOpen
  \bibfield  {author} {\bibinfo {author} {\bibfnamefont {A.~G.}\ \bibnamefont {Hart}},\ }\bibfield  {title} {\bibinfo {title} {Generalised synchronisations, embeddings, and approximations for continuous time reservoir computers},\ }\href@noop {} {\bibfield  {journal} {\bibinfo  {journal} {Physica D: Nonlinear Phenomena}\ }\textbf {\bibinfo {volume} {458}},\ \bibinfo {pages} {133956} (\bibinfo {year} {2024}{\natexlab{a}})}\BibitemShut {NoStop}%
\bibitem [{\citenamefont {Hart}(2024{\natexlab{b}})}]{hart2024attractor}%
  \BibitemOpen
  \bibfield  {author} {\bibinfo {author} {\bibfnamefont {J.~D.}\ \bibnamefont {Hart}},\ }\bibfield  {title} {\bibinfo {title} {Attractor reconstruction with reservoir computers: The effect of the reservoir’s conditional lyapunov exponents on faithful attractor reconstruction},\ }\href@noop {} {\bibfield  {journal} {\bibinfo  {journal} {Chaos: An Interdisciplinary Journal of Nonlinear Science}\ }\textbf {\bibinfo {volume} {34}} (\bibinfo {year} {2024}{\natexlab{b}})}\BibitemShut {NoStop}%
\bibitem [{\citenamefont {Hess}\ \emph {et~al.}(2023)\citenamefont {Hess}, \citenamefont {Monfared}, \citenamefont {Brenner},\ and\ \citenamefont {Durstewitz}}]{hess2023generalized}%
  \BibitemOpen
  \bibfield  {author} {\bibinfo {author} {\bibfnamefont {F.}~\bibnamefont {Hess}}, \bibinfo {author} {\bibfnamefont {Z.}~\bibnamefont {Monfared}}, \bibinfo {author} {\bibfnamefont {M.}~\bibnamefont {Brenner}},\ and\ \bibinfo {author} {\bibfnamefont {D.}~\bibnamefont {Durstewitz}},\ }\bibfield  {title} {\bibinfo {title} {Generalized teacher forcing for learning chaotic dynamics},\ }\href@noop {} {\bibfield  {journal} {\bibinfo  {journal} {arXiv preprint arXiv:2306.04406}\ } (\bibinfo {year} {2023})}\BibitemShut {NoStop}%
\bibitem [{\citenamefont {Stephens}\ \emph {et~al.}(2008)\citenamefont {Stephens}, \citenamefont {Johnson-Kerner}, \citenamefont {Bialek},\ and\ \citenamefont {Ryu}}]{stephens2008dimensionality}%
  \BibitemOpen
  \bibfield  {author} {\bibinfo {author} {\bibfnamefont {G.~J.}\ \bibnamefont {Stephens}}, \bibinfo {author} {\bibfnamefont {B.}~\bibnamefont {Johnson-Kerner}}, \bibinfo {author} {\bibfnamefont {W.}~\bibnamefont {Bialek}},\ and\ \bibinfo {author} {\bibfnamefont {W.~S.}\ \bibnamefont {Ryu}},\ }\bibfield  {title} {\bibinfo {title} {Dimensionality and dynamics in the behavior of c. elegans},\ }\href@noop {} {\bibfield  {journal} {\bibinfo  {journal} {PLoS computational biology}\ }\textbf {\bibinfo {volume} {4}},\ \bibinfo {pages} {e1000028} (\bibinfo {year} {2008})}\BibitemShut {NoStop}%
\bibitem [{\citenamefont {May}(1976)}]{may1976}%
  \BibitemOpen
  \bibfield  {author} {\bibinfo {author} {\bibfnamefont {R.~M.}\ \bibnamefont {May}},\ }\bibfield  {title} {\bibinfo {title} {Simple mathematical models with very complicated dynamics},\ }\href@noop {} {\bibfield  {journal} {\bibinfo  {journal} {Nature}\ }\textbf {\bibinfo {volume} {261}},\ \bibinfo {pages} {459} (\bibinfo {year} {1976})}\BibitemShut {NoStop}%
\bibitem [{\citenamefont {Platt}\ \emph {et~al.}(2023)\citenamefont {Platt}, \citenamefont {Penny}, \citenamefont {Smith}, \citenamefont {Chen},\ and\ \citenamefont {Abarbanel}}]{platt_constraining_2023}%
  \BibitemOpen
  \bibfield  {author} {\bibinfo {author} {\bibfnamefont {J.~A.}\ \bibnamefont {Platt}}, \bibinfo {author} {\bibfnamefont {S.~G.}\ \bibnamefont {Penny}}, \bibinfo {author} {\bibfnamefont {T.~A.}\ \bibnamefont {Smith}}, \bibinfo {author} {\bibfnamefont {T.-C.}\ \bibnamefont {Chen}},\ and\ \bibinfo {author} {\bibfnamefont {H.~D.~I.}\ \bibnamefont {Abarbanel}},\ }\bibfield  {title} {\bibinfo {title} {Constraining chaos: {Enforcing} dynamical invariants in the training of reservoir computers},\ }\href {https://doi.org/10.1063/5.0156999} {\bibfield  {journal} {\bibinfo  {journal} {Chaos: An Interdisciplinary Journal of Nonlinear Science}\ }\textbf {\bibinfo {volume} {33}},\ \bibinfo {pages} {103107} (\bibinfo {year} {2023})}\BibitemShut {NoStop}%
\bibitem [{\citenamefont {Massey}(1951)}]{massey_kolmogorov-smirnov_1951}%
  \BibitemOpen
  \bibfield  {author} {\bibinfo {author} {\bibfnamefont {F.~J.}\ \bibnamefont {Massey}},\ }\bibfield  {title} {\bibinfo {title} {The {Kolmogorov}-{Smirnov} {Test} for {Goodness} of {Fit}},\ }\href {https://doi.org/10.2307/2280095} {\bibfield  {journal} {\bibinfo  {journal} {Journal of the American Statistical Association}\ }\textbf {\bibinfo {volume} {46}},\ \bibinfo {pages} {68} (\bibinfo {year} {1951})},\ \bibinfo {note} {publisher: [American Statistical Association, Taylor \& Francis, Ltd.]}\BibitemShut {NoStop}%
\bibitem [{\citenamefont {Chirikov}(1979)}]{Chirikov1979}%
  \BibitemOpen
  \bibfield  {author} {\bibinfo {author} {\bibfnamefont {B.}~\bibnamefont {Chirikov}},\ }\bibfield  {title} {\bibinfo {title} {A universal instability of many-dimensional oscillator systems},\ }\href {https://doi.org/https://doi.org/10.1016/0370-1573(79)90023-1} {\bibfield  {journal} {\bibinfo  {journal} {Physics Reports}\ }\textbf {\bibinfo {volume} {52}},\ \bibinfo {pages} {263} (\bibinfo {year} {1979})}\BibitemShut {NoStop}%
\bibitem [{\citenamefont {Takens}(1981)}]{citeulike:2735031}%
  \BibitemOpen
  \bibfield  {author} {\bibinfo {author} {\bibfnamefont {F.}~\bibnamefont {Takens}},\ }\bibfield  {title} {\bibinfo {title} {{Detecting Strange Attractors in Turbulence}},\ }in\ \href {https://doi.org/10.1007/bfb0091924} {\emph {\bibinfo {booktitle} {Dynamical Systems and Turbulence, Warwick 1980}}},\ \bibinfo {series} {Lecture Notes in Mathematics}, Vol.\ \bibinfo {volume} {898},\ \bibinfo {editor} {edited by\ \bibinfo {editor} {\bibfnamefont {D.}~\bibnamefont {Rand}}\ and\ \bibinfo {editor} {\bibfnamefont {L.-S.}\ \bibnamefont {Young}}}\ (\bibinfo  {publisher} {Springer},\ \bibinfo {address} {Berlin},\ \bibinfo {year} {1981})\ Chap.~\bibinfo {chapter} {21}, pp.\ \bibinfo {pages} {366--381}\BibitemShut {NoStop}%
\bibitem [{\citenamefont {Zhang}\ \emph {et~al.}(2021)\citenamefont {Zhang}, \citenamefont {Fan}, \citenamefont {Wang},\ and\ \citenamefont {Wang}}]{zhang2021learning}%
  \BibitemOpen
  \bibfield  {author} {\bibinfo {author} {\bibfnamefont {H.}~\bibnamefont {Zhang}}, \bibinfo {author} {\bibfnamefont {H.}~\bibnamefont {Fan}}, \bibinfo {author} {\bibfnamefont {L.}~\bibnamefont {Wang}},\ and\ \bibinfo {author} {\bibfnamefont {X.}~\bibnamefont {Wang}},\ }\bibfield  {title} {\bibinfo {title} {Learning hamiltonian dynamics with reservoir computing},\ }\href@noop {} {\bibfield  {journal} {\bibinfo  {journal} {Physical Review E}\ }\textbf {\bibinfo {volume} {104}},\ \bibinfo {pages} {024205} (\bibinfo {year} {2021})}\BibitemShut {NoStop}%
\bibitem [{\citenamefont {Jaeger}\ \emph {et~al.}(2007)\citenamefont {Jaeger}, \citenamefont {Luko{\v{s}}evi{\v{c}}ius}, \citenamefont {Popovici},\ and\ \citenamefont {Siewert}}]{jaeger2007optimization}%
  \BibitemOpen
  \bibfield  {author} {\bibinfo {author} {\bibfnamefont {H.}~\bibnamefont {Jaeger}}, \bibinfo {author} {\bibfnamefont {M.}~\bibnamefont {Luko{\v{s}}evi{\v{c}}ius}}, \bibinfo {author} {\bibfnamefont {D.}~\bibnamefont {Popovici}},\ and\ \bibinfo {author} {\bibfnamefont {U.}~\bibnamefont {Siewert}},\ }\bibfield  {title} {\bibinfo {title} {Optimization and applications of echo state networks with leaky-integrator neurons},\ }\href@noop {} {\bibfield  {journal} {\bibinfo  {journal} {Neural networks}\ }\textbf {\bibinfo {volume} {20}},\ \bibinfo {pages} {335} (\bibinfo {year} {2007})}\BibitemShut {NoStop}%
\bibitem [{\citenamefont {Ebato}\ \emph {et~al.}(2024)\citenamefont {Ebato}, \citenamefont {Nobukawa}, \citenamefont {Sakemi}, \citenamefont {Nishimura}, \citenamefont {Kanamaru}, \citenamefont {Sviridova},\ and\ \citenamefont {Aihara}}]{ebato2024impact}%
  \BibitemOpen
  \bibfield  {author} {\bibinfo {author} {\bibfnamefont {Y.}~\bibnamefont {Ebato}}, \bibinfo {author} {\bibfnamefont {S.}~\bibnamefont {Nobukawa}}, \bibinfo {author} {\bibfnamefont {Y.}~\bibnamefont {Sakemi}}, \bibinfo {author} {\bibfnamefont {H.}~\bibnamefont {Nishimura}}, \bibinfo {author} {\bibfnamefont {T.}~\bibnamefont {Kanamaru}}, \bibinfo {author} {\bibfnamefont {N.}~\bibnamefont {Sviridova}},\ and\ \bibinfo {author} {\bibfnamefont {K.}~\bibnamefont {Aihara}},\ }\bibfield  {title} {\bibinfo {title} {Impact of time-history terms on reservoir dynamics and prediction accuracy in echo state networks},\ }\href@noop {} {\bibfield  {journal} {\bibinfo  {journal} {Scientific Reports}\ }\textbf {\bibinfo {volume} {14}},\ \bibinfo {pages} {8631} (\bibinfo {year} {2024})}\BibitemShut {NoStop}%
\bibitem [{\citenamefont {Bonneel}\ \emph {et~al.}(2015)\citenamefont {Bonneel}, \citenamefont {Rabin}, \citenamefont {Peyr{\'e}},\ and\ \citenamefont {Pfister}}]{bonneel2015sliced}%
  \BibitemOpen
  \bibfield  {author} {\bibinfo {author} {\bibfnamefont {N.}~\bibnamefont {Bonneel}}, \bibinfo {author} {\bibfnamefont {J.}~\bibnamefont {Rabin}}, \bibinfo {author} {\bibfnamefont {G.}~\bibnamefont {Peyr{\'e}}},\ and\ \bibinfo {author} {\bibfnamefont {H.}~\bibnamefont {Pfister}},\ }\bibfield  {title} {\bibinfo {title} {Sliced and radon wasserstein barycenters of measures},\ }\href@noop {} {\bibfield  {journal} {\bibinfo  {journal} {Journal of Mathematical Imaging and Vision}\ }\textbf {\bibinfo {volume} {51}},\ \bibinfo {pages} {22} (\bibinfo {year} {2015})}\BibitemShut {NoStop}%
\bibitem [{\citenamefont {Zhen}\ and\ \citenamefont {Samuel}(2015)}]{zhen2015}%
  \BibitemOpen
  \bibfield  {author} {\bibinfo {author} {\bibfnamefont {M.}~\bibnamefont {Zhen}}\ and\ \bibinfo {author} {\bibfnamefont {A.~D.}\ \bibnamefont {Samuel}},\ }\bibfield  {title} {\bibinfo {title} {C. elegans locomotion: small circuits, complex functions},\ }\href@noop {} {\bibfield  {journal} {\bibinfo  {journal} {Current opinion in neurobiology}\ }\textbf {\bibinfo {volume} {33}},\ \bibinfo {pages} {117} (\bibinfo {year} {2015})}\BibitemShut {NoStop}%
\bibitem [{\citenamefont {Hallinen}\ \emph {et~al.}(2021)\citenamefont {Hallinen}, \citenamefont {Dempsey}, \citenamefont {Scholz}, \citenamefont {Yu}, \citenamefont {Linder}, \citenamefont {Randi}, \citenamefont {Sharma}, \citenamefont {Shaevitz},\ and\ \citenamefont {Leifer}}]{hallinen2021}%
  \BibitemOpen
  \bibfield  {author} {\bibinfo {author} {\bibfnamefont {K.~M.}\ \bibnamefont {Hallinen}}, \bibinfo {author} {\bibfnamefont {R.}~\bibnamefont {Dempsey}}, \bibinfo {author} {\bibfnamefont {M.}~\bibnamefont {Scholz}}, \bibinfo {author} {\bibfnamefont {X.}~\bibnamefont {Yu}}, \bibinfo {author} {\bibfnamefont {A.}~\bibnamefont {Linder}}, \bibinfo {author} {\bibfnamefont {F.}~\bibnamefont {Randi}}, \bibinfo {author} {\bibfnamefont {A.~K.}\ \bibnamefont {Sharma}}, \bibinfo {author} {\bibfnamefont {J.~W.}\ \bibnamefont {Shaevitz}},\ and\ \bibinfo {author} {\bibfnamefont {A.~M.}\ \bibnamefont {Leifer}},\ }\bibfield  {title} {\bibinfo {title} {Decoding locomotion from population neural activity in moving \textit{C. elegans}},\ }\href {https://doi.org/10.7554/eLife.66135} {\bibfield  {journal} {\bibinfo  {journal} {eLife}\ }\textbf {\bibinfo {volume} {10}},\ \bibinfo {pages} {e66135} (\bibinfo {year} {2021})}\BibitemShut {NoStop}%
\bibitem [{\citenamefont {Atanas}\ \emph {et~al.}(2023)\citenamefont {Atanas}, \citenamefont {Kim}, \citenamefont {Wang}, \citenamefont {Bueno}, \citenamefont {Becker}, \citenamefont {Kang}, \citenamefont {Park}, \citenamefont {Kramer}, \citenamefont {Wan}, \citenamefont {Baskoylu} \emph {et~al.}}]{atanas2023}%
  \BibitemOpen
  \bibfield  {author} {\bibinfo {author} {\bibfnamefont {A.~A.}\ \bibnamefont {Atanas}}, \bibinfo {author} {\bibfnamefont {J.}~\bibnamefont {Kim}}, \bibinfo {author} {\bibfnamefont {Z.}~\bibnamefont {Wang}}, \bibinfo {author} {\bibfnamefont {E.}~\bibnamefont {Bueno}}, \bibinfo {author} {\bibfnamefont {M.}~\bibnamefont {Becker}}, \bibinfo {author} {\bibfnamefont {D.}~\bibnamefont {Kang}}, \bibinfo {author} {\bibfnamefont {J.}~\bibnamefont {Park}}, \bibinfo {author} {\bibfnamefont {T.~S.}\ \bibnamefont {Kramer}}, \bibinfo {author} {\bibfnamefont {F.~K.}\ \bibnamefont {Wan}}, \bibinfo {author} {\bibfnamefont {S.}~\bibnamefont {Baskoylu}}, \emph {et~al.},\ }\bibfield  {title} {\bibinfo {title} {Brain-wide representations of behavior spanning multiple timescales and states in c. elegans},\ }\href@noop {} {\bibfield  {journal} {\bibinfo  {journal} {Cell}\ }\textbf {\bibinfo {volume} {186}},\ \bibinfo {pages} {4134} (\bibinfo {year} {2023})}\BibitemShut {NoStop}%
\bibitem [{\citenamefont {Bryant}\ \emph {et~al.}(1990)\citenamefont {Bryant}, \citenamefont {Brown},\ and\ \citenamefont {Abarbanel}}]{bryant1990lyapunov}%
  \BibitemOpen
  \bibfield  {author} {\bibinfo {author} {\bibfnamefont {P.}~\bibnamefont {Bryant}}, \bibinfo {author} {\bibfnamefont {R.}~\bibnamefont {Brown}},\ and\ \bibinfo {author} {\bibfnamefont {H.~D.}\ \bibnamefont {Abarbanel}},\ }\bibfield  {title} {\bibinfo {title} {Lyapunov exponents from observed time series},\ }\href@noop {} {\bibfield  {journal} {\bibinfo  {journal} {Physical Review Letters}\ }\textbf {\bibinfo {volume} {65}},\ \bibinfo {pages} {1523} (\bibinfo {year} {1990})}\BibitemShut {NoStop}%
\bibitem [{\citenamefont {Dingle}\ \emph {et~al.}(2018)\citenamefont {Dingle}, \citenamefont {Camargo},\ and\ \citenamefont {Louis}}]{dingle2018}%
  \BibitemOpen
  \bibfield  {author} {\bibinfo {author} {\bibfnamefont {K.}~\bibnamefont {Dingle}}, \bibinfo {author} {\bibfnamefont {C.~Q.}\ \bibnamefont {Camargo}},\ and\ \bibinfo {author} {\bibfnamefont {A.~A.}\ \bibnamefont {Louis}},\ }\bibfield  {title} {\bibinfo {title} {Input--output maps are strongly biased towards simple outputs},\ }\href@noop {} {\bibfield  {journal} {\bibinfo  {journal} {Nature communications}\ }\textbf {\bibinfo {volume} {9}},\ \bibinfo {pages} {761} (\bibinfo {year} {2018})}\BibitemShut {NoStop}%
\bibitem [{\citenamefont {Dingle}\ \emph {et~al.}(2024)\citenamefont {Dingle}, \citenamefont {Alaskandarani}, \citenamefont {Hamzi},\ and\ \citenamefont {Louis}}]{dingle2024}%
  \BibitemOpen
  \bibfield  {author} {\bibinfo {author} {\bibfnamefont {K.}~\bibnamefont {Dingle}}, \bibinfo {author} {\bibfnamefont {M.}~\bibnamefont {Alaskandarani}}, \bibinfo {author} {\bibfnamefont {B.}~\bibnamefont {Hamzi}},\ and\ \bibinfo {author} {\bibfnamefont {A.~A.}\ \bibnamefont {Louis}},\ }\bibfield  {title} {\bibinfo {title} {Exploring simplicity bias in 1d dynamical systems},\ }\href@noop {} {\bibfield  {journal} {\bibinfo  {journal} {Entropy}\ }\textbf {\bibinfo {volume} {26}},\ \bibinfo {pages} {426} (\bibinfo {year} {2024})}\BibitemShut {NoStop}%
\bibitem [{\citenamefont {Brudno}(1982)}]{brudno1982}%
  \BibitemOpen
  \bibfield  {author} {\bibinfo {author} {\bibfnamefont {A.}~\bibnamefont {Brudno}},\ }\bibfield  {title} {\bibinfo {title} {Entropy and the complexity of the trajectories of a dynamic system},\ }\href@noop {} {\bibfield  {journal} {\bibinfo  {journal} {Trudy Moskovskogo Matematicheskogo Obshchestva}\ }\textbf {\bibinfo {volume} {44}},\ \bibinfo {pages} {124} (\bibinfo {year} {1982})}\BibitemShut {NoStop}%
\bibitem [{\citenamefont {Pesin}(1977)}]{pesin1977characteristic}%
  \BibitemOpen
  \bibfield  {author} {\bibinfo {author} {\bibfnamefont {Y.~B.}\ \bibnamefont {Pesin}},\ }\bibfield  {title} {\bibinfo {title} {Characteristic lyapunov exponents and smooth ergodic theory},\ }\href@noop {} {\bibfield  {journal} {\bibinfo  {journal} {Russian Mathematical Surveys}\ }\textbf {\bibinfo {volume} {32}},\ \bibinfo {pages} {55} (\bibinfo {year} {1977})}\BibitemShut {NoStop}%
\bibitem [{\citenamefont {Kantz}\ \emph {et~al.}(2013)\citenamefont {Kantz}, \citenamefont {Radons},\ and\ \citenamefont {Yang}}]{kantz2013}%
  \BibitemOpen
  \bibfield  {author} {\bibinfo {author} {\bibfnamefont {H.}~\bibnamefont {Kantz}}, \bibinfo {author} {\bibfnamefont {G.}~\bibnamefont {Radons}},\ and\ \bibinfo {author} {\bibfnamefont {H.}~\bibnamefont {Yang}},\ }\bibfield  {title} {\bibinfo {title} {The problem of spurious lyapunov exponents in time series analysis and its solution by covariant lyapunov vectors},\ }\href@noop {} {\bibfield  {journal} {\bibinfo  {journal} {Journal of Physics A: Mathematical and Theoretical}\ }\textbf {\bibinfo {volume} {46}},\ \bibinfo {pages} {254009} (\bibinfo {year} {2013})}\BibitemShut {NoStop}%
\bibitem [{\citenamefont {Duarte}\ and\ \citenamefont {Eisencraft}(2024)}]{duarte2024denoising}%
  \BibitemOpen
  \bibfield  {author} {\bibinfo {author} {\bibfnamefont {A.~L.}\ \bibnamefont {Duarte}}\ and\ \bibinfo {author} {\bibfnamefont {M.}~\bibnamefont {Eisencraft}},\ }\bibfield  {title} {\bibinfo {title} {Denoising of discrete-time chaotic signals using echo state networks},\ }\href@noop {} {\bibfield  {journal} {\bibinfo  {journal} {Signal Processing}\ }\textbf {\bibinfo {volume} {214}},\ \bibinfo {pages} {109252} (\bibinfo {year} {2024})}\BibitemShut {NoStop}%
\bibitem [{\citenamefont {Huang}\ \emph {et~al.}(2020)\citenamefont {Huang}, \citenamefont {Fu},\ and\ \citenamefont {Franzke}}]{huang_detecting_2020}%
  \BibitemOpen
  \bibfield  {author} {\bibinfo {author} {\bibfnamefont {Y.}~\bibnamefont {Huang}}, \bibinfo {author} {\bibfnamefont {Z.}~\bibnamefont {Fu}},\ and\ \bibinfo {author} {\bibfnamefont {C.~L.~E.}\ \bibnamefont {Franzke}},\ }\bibfield  {title} {\bibinfo {title} {Detecting causality from time series in a machine learning framework},\ }\href {https://doi.org/10.1063/5.0007670} {\bibfield  {journal} {\bibinfo  {journal} {Chaos: An Interdisciplinary Journal of Nonlinear Science}\ }\textbf {\bibinfo {volume} {30}},\ \bibinfo {pages} {063116} (\bibinfo {year} {2020})}\BibitemShut {NoStop}%
\bibitem [{\citenamefont {Sugihara}\ \emph {et~al.}(2012)\citenamefont {Sugihara}, \citenamefont {May}, \citenamefont {Ye}, \citenamefont {Hsieh}, \citenamefont {Deyle}, \citenamefont {Fogarty},\ and\ \citenamefont {Munch}}]{sugihara_detecting_2012}%
  \BibitemOpen
  \bibfield  {author} {\bibinfo {author} {\bibfnamefont {G.}~\bibnamefont {Sugihara}}, \bibinfo {author} {\bibfnamefont {R.}~\bibnamefont {May}}, \bibinfo {author} {\bibfnamefont {H.}~\bibnamefont {Ye}}, \bibinfo {author} {\bibfnamefont {C.-h.}\ \bibnamefont {Hsieh}}, \bibinfo {author} {\bibfnamefont {E.}~\bibnamefont {Deyle}}, \bibinfo {author} {\bibfnamefont {M.}~\bibnamefont {Fogarty}},\ and\ \bibinfo {author} {\bibfnamefont {S.}~\bibnamefont {Munch}},\ }\bibfield  {title} {\bibinfo {title} {Detecting {Causality} in {Complex} {Ecosystems}},\ }\href {https://doi.org/10.1126/science.1227079} {\bibfield  {journal} {\bibinfo  {journal} {Science}\ }\textbf {\bibinfo {volume} {338}},\ \bibinfo {pages} {496} (\bibinfo {year} {2012})}\BibitemShut {NoStop}%
\bibitem [{\citenamefont {Krishnamurthy}\ \emph {et~al.}(2022)\citenamefont {Krishnamurthy}, \citenamefont {Can},\ and\ \citenamefont {Schwab}}]{krishnamurthy2022}%
  \BibitemOpen
  \bibfield  {author} {\bibinfo {author} {\bibfnamefont {K.}~\bibnamefont {Krishnamurthy}}, \bibinfo {author} {\bibfnamefont {T.}~\bibnamefont {Can}},\ and\ \bibinfo {author} {\bibfnamefont {D.~J.}\ \bibnamefont {Schwab}},\ }\bibfield  {title} {\bibinfo {title} {Theory of gating in recurrent neural networks},\ }\href@noop {} {\bibfield  {journal} {\bibinfo  {journal} {Physical Review X}\ }\textbf {\bibinfo {volume} {12}},\ \bibinfo {pages} {011011} (\bibinfo {year} {2022})}\BibitemShut {NoStop}%
\bibitem [{\citenamefont {Benettin}\ \emph {et~al.}(1980)\citenamefont {Benettin}, \citenamefont {Galgani}, \citenamefont {Giorgilli},\ and\ \citenamefont {Strelcyn}}]{benettin1980lyapunov}%
  \BibitemOpen
  \bibfield  {author} {\bibinfo {author} {\bibfnamefont {G.}~\bibnamefont {Benettin}}, \bibinfo {author} {\bibfnamefont {L.}~\bibnamefont {Galgani}}, \bibinfo {author} {\bibfnamefont {A.}~\bibnamefont {Giorgilli}},\ and\ \bibinfo {author} {\bibfnamefont {J.-M.}\ \bibnamefont {Strelcyn}},\ }\bibfield  {title} {\bibinfo {title} {Lyapunov characteristic exponents for smooth dynamical systems and for hamiltonian systems; a method for computing all of them. part 1: Theory},\ }\href@noop {} {\bibfield  {journal} {\bibinfo  {journal} {Meccanica}\ }\textbf {\bibinfo {volume} {15}},\ \bibinfo {pages} {9} (\bibinfo {year} {1980})}\BibitemShut {NoStop}%
\bibitem [{\citenamefont {Eckmann}\ and\ \citenamefont {Ruelle}(1985)}]{eckmann1985ergodic}%
  \BibitemOpen
  \bibfield  {author} {\bibinfo {author} {\bibfnamefont {J.-P.}\ \bibnamefont {Eckmann}}\ and\ \bibinfo {author} {\bibfnamefont {D.}~\bibnamefont {Ruelle}},\ }\bibfield  {title} {\bibinfo {title} {Ergodic theory of chaos and strange attractors},\ }\href@noop {} {\bibfield  {journal} {\bibinfo  {journal} {Reviews of modern physics}\ }\textbf {\bibinfo {volume} {57}},\ \bibinfo {pages} {617} (\bibinfo {year} {1985})}\BibitemShut {NoStop}%
\bibitem [{\citenamefont {Hodges~Jr}(1958)}]{hodges1958significance}%
  \BibitemOpen
  \bibfield  {author} {\bibinfo {author} {\bibfnamefont {J.}~\bibnamefont {Hodges~Jr}},\ }\bibfield  {title} {\bibinfo {title} {The significance probability of the smirnov two-sample test},\ }\href@noop {} {\bibfield  {journal} {\bibinfo  {journal} {Arkiv f{\"o}r matematik}\ }\textbf {\bibinfo {volume} {3}},\ \bibinfo {pages} {469} (\bibinfo {year} {1958})}\BibitemShut {NoStop}%
\bibitem [{\citenamefont {Hairer}\ \emph {et~al.}(1993)\citenamefont {Hairer}, \citenamefont {Wanner},\ and\ \citenamefont {N{\o}rsett}}]{hairer1993solving}%
  \BibitemOpen
  \bibfield  {author} {\bibinfo {author} {\bibfnamefont {E.}~\bibnamefont {Hairer}}, \bibinfo {author} {\bibfnamefont {G.}~\bibnamefont {Wanner}},\ and\ \bibinfo {author} {\bibfnamefont {S.~P.}\ \bibnamefont {N{\o}rsett}},\ }\href@noop {} {\emph {\bibinfo {title} {Solving ordinary differential equations I: Nonstiff problems}}}\ (\bibinfo  {publisher} {Springer},\ \bibinfo {year} {1993})\BibitemShut {NoStop}%
\bibitem [{\citenamefont {Bickel}\ and\ \citenamefont {Sakov}(2008)}]{bickel2008choice}%
  \BibitemOpen
  \bibfield  {author} {\bibinfo {author} {\bibfnamefont {P.~J.}\ \bibnamefont {Bickel}}\ and\ \bibinfo {author} {\bibfnamefont {A.}~\bibnamefont {Sakov}},\ }\bibfield  {title} {\bibinfo {title} {On the choice of m in the m out of n bootstrap and confidence bounds for extrema},\ }\href@noop {} {\bibfield  {journal} {\bibinfo  {journal} {Statistica Sinica}\ ,\ \bibinfo {pages} {967}} (\bibinfo {year} {2008})}\BibitemShut {NoStop}%
\end{thebibliography}%

\newpage
\section*{Supplementary Material}
\renewcommand{\theequation}{S\arabic{equation}}
\renewcommand{\thefigure}{S\arabic{figure}}
\setcounter{figure}{0}

\begin{figure}[h]
\begin{center} 
\includegraphics[width=0.9\linewidth]{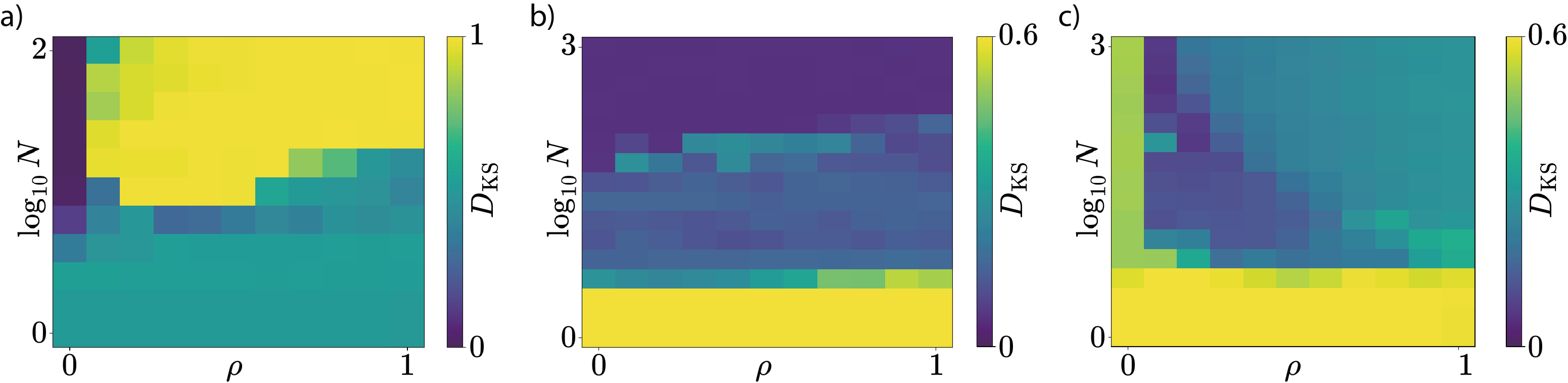}
\caption{{\bf Optimization of reservoir computing by minimizing the Kolmogorov-Smirnov distance.} 
(a) Distribution reconstruction errors measured by the Kolmogorov-Smirnov distance, $D_\mathrm{KS}$, between the true and predicted invariant distributions of the logistic map. Errors are shown across different numbers of neurons $N$ and spectral radii of the connection matrix $\rho$. For each parameter pair, we calculated the median of $D_\mathrm{KS}$ across 30 network realizations. 
(b) Reconstruction errors measured by $D_\mathrm{KS}$ for networks modeling the standard map from partial observation $\{\theta\}$. 
(c) Reconstruction errors measured by $D_\mathrm{KS}$ for networks modeling the standard map from full observation $\{\theta, p\}$. 
}
\label{fig:d_ks_opt}
\end{center}
\end{figure}

\begin{figure}[h]
\begin{center} 
\includegraphics[width=0.5\linewidth]{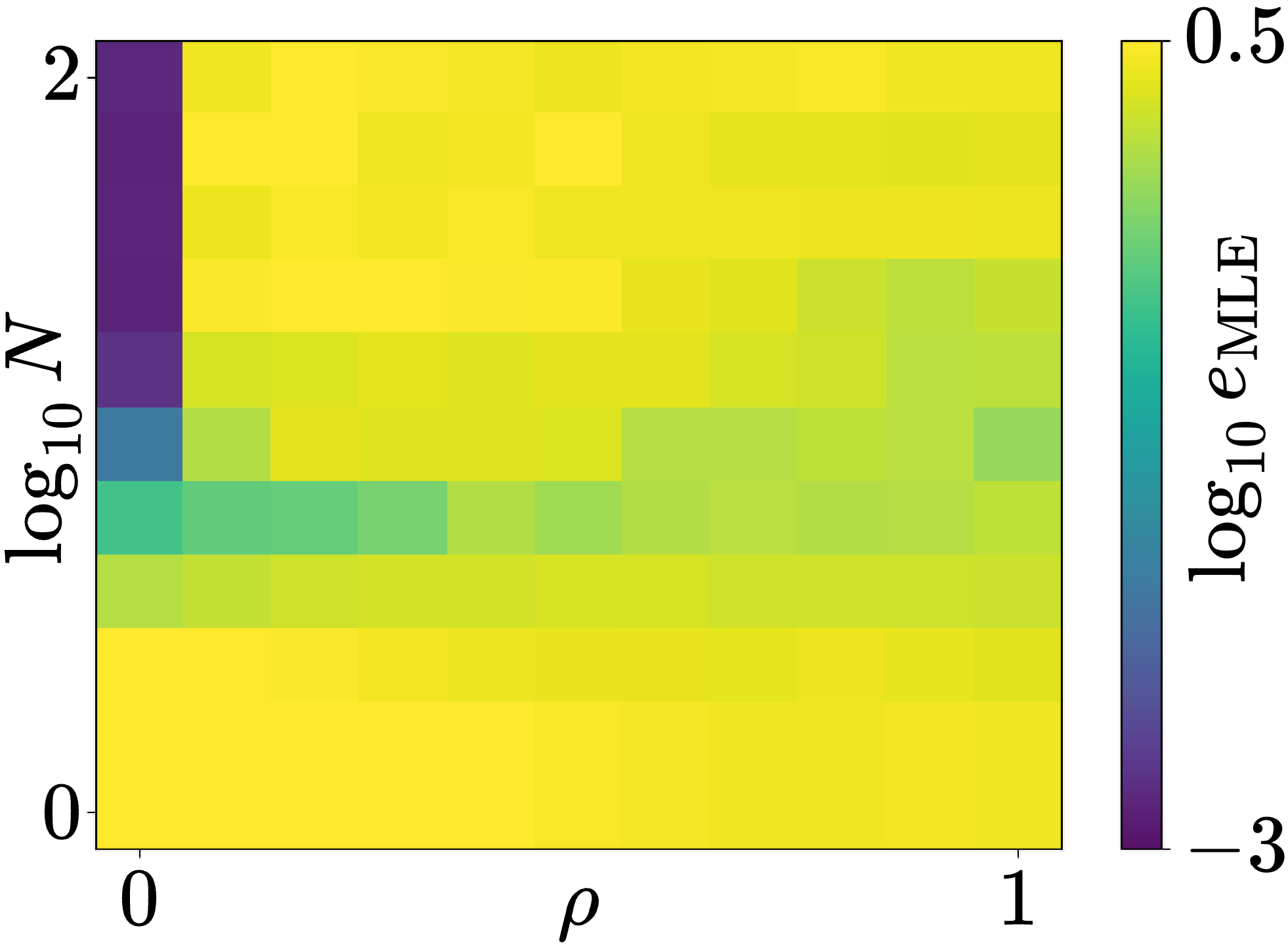}
\caption{{\bf Lyapunov exponent estimation error from our reservoir computing approach applied to the logistic map.} We evaluate the performance of Lyapunov exponent estimation from reservoir networks with a different number of neurons $N$ and a spectral radius of the connection matrix $\rho$ by measuring the absolute relative error of the maximal Lyapunov exponent (MLE), $e_\mathrm{MLE} = \left|\frac{\lambda^\mathrm{estimate}_\mathrm{max}-\lambda^\mathrm{true}_\mathrm{max}}{\lambda^\mathrm{true}_\mathrm{max}}\right|$. For each pair of $N$ and $\rho$, we calculate the median of $\log_{10}e_\mathrm{MLE}$ from 30 realizations of the network. Since we fully observe this 1D system, no reservoir memory is required and the best estimates come from zero spectral radius. 
}
\label{fig:mle_err_logistic}
\end{center}
\end{figure}

\begin{figure}[h]
\begin{center} 
\includegraphics[width=0.5\linewidth]{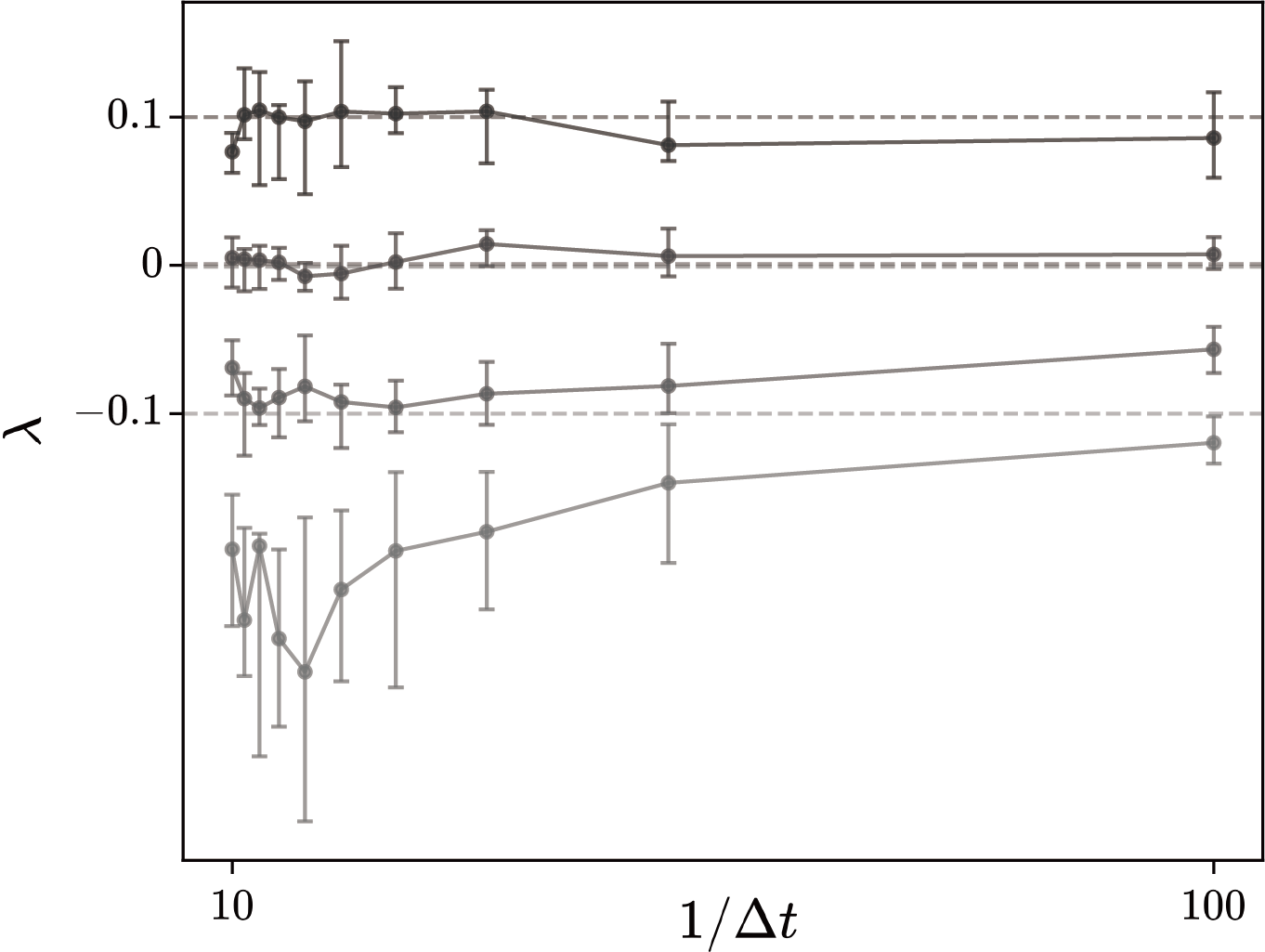}
\caption{{\bf Convergence of the Lyapunov spectrum with increasing temporal resolution in our reservoir analysis of the (partially observed) double pendulum.} We show the first four Lyapunov exponents of the autonomous reservoir as a function of sampling frequency $1/\Delta t$. Horizontal dashed lines indicate the analytical Lyapunov exponents. Error bars represent 95\% confidence intervals bootstrapped from $10^4$ samples.}
\label{fig:dt_convergence}
\end{center}
\end{figure}

\begin{figure}[h]
\begin{center} 
\includegraphics[width=0.55\linewidth]{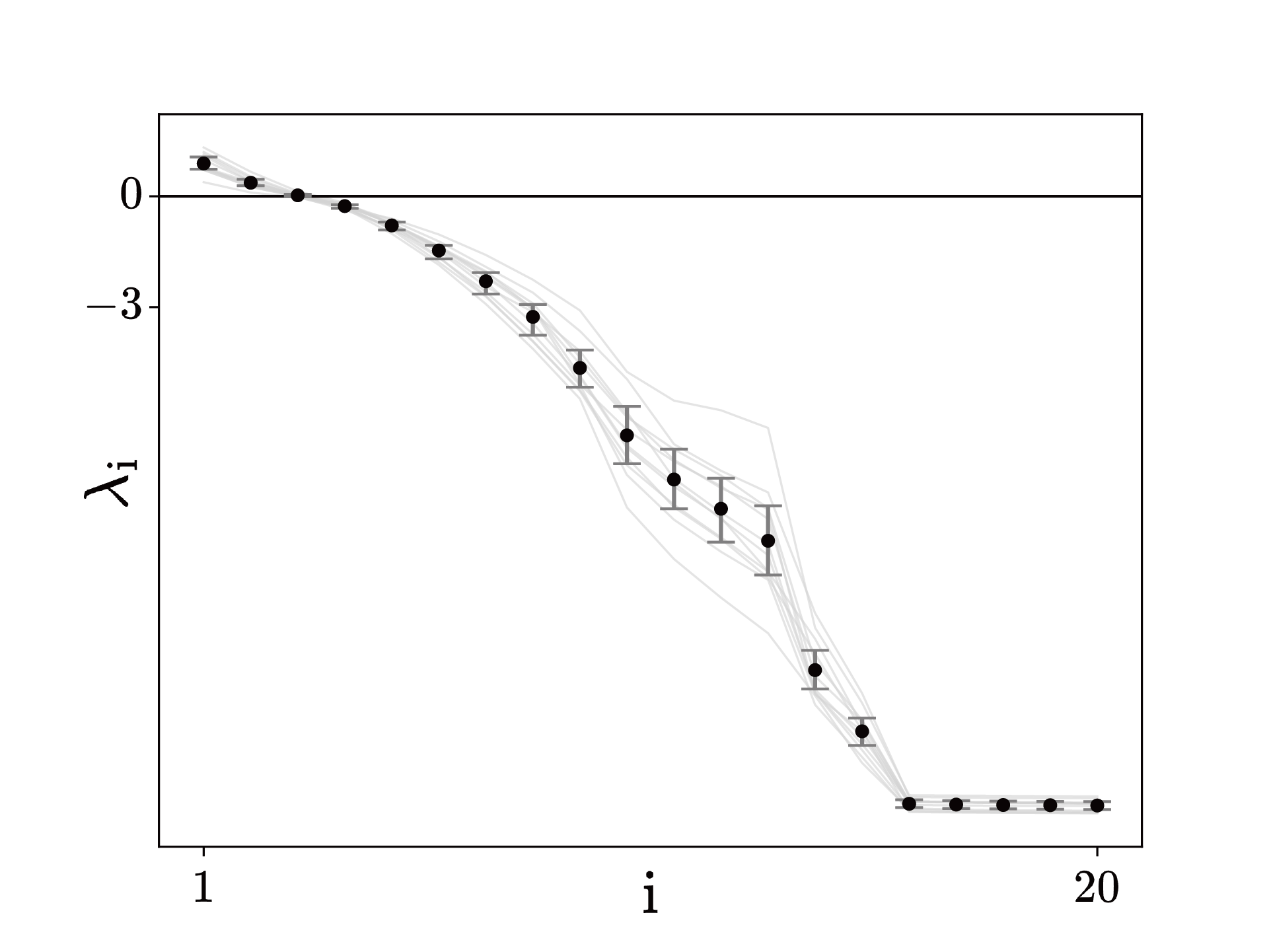}
\caption{{\bf Ensemble estimate of the Lyapunov spectrum across $N=12$ foraging {\em C. elegans}}. For each worm we trained $M=100$ ESNs at the optimised hyperparameters and selected the bootstrap-best reservoir via $m$-out-of-$M$ subsample bootstrap ($m=40$, without replacement). For each bootstrap-winning ESN, we computed the Lyapunov spectrum from $N_{\rm traj}=15$ independent trajectory realisations. Each iteration resamples 12 worms with replacement, draws one spectrum at random from each selected worm's bootstrap pool, and reports the mean across the 12. Error bars are the 2.5 - 97.5\% percentile interval across the 5000 iterations.}
\label{fig:S_all}
\end{center}
\end{figure}
\begin{figure}[h]
\begin{center} 
\includegraphics[width=1\linewidth]{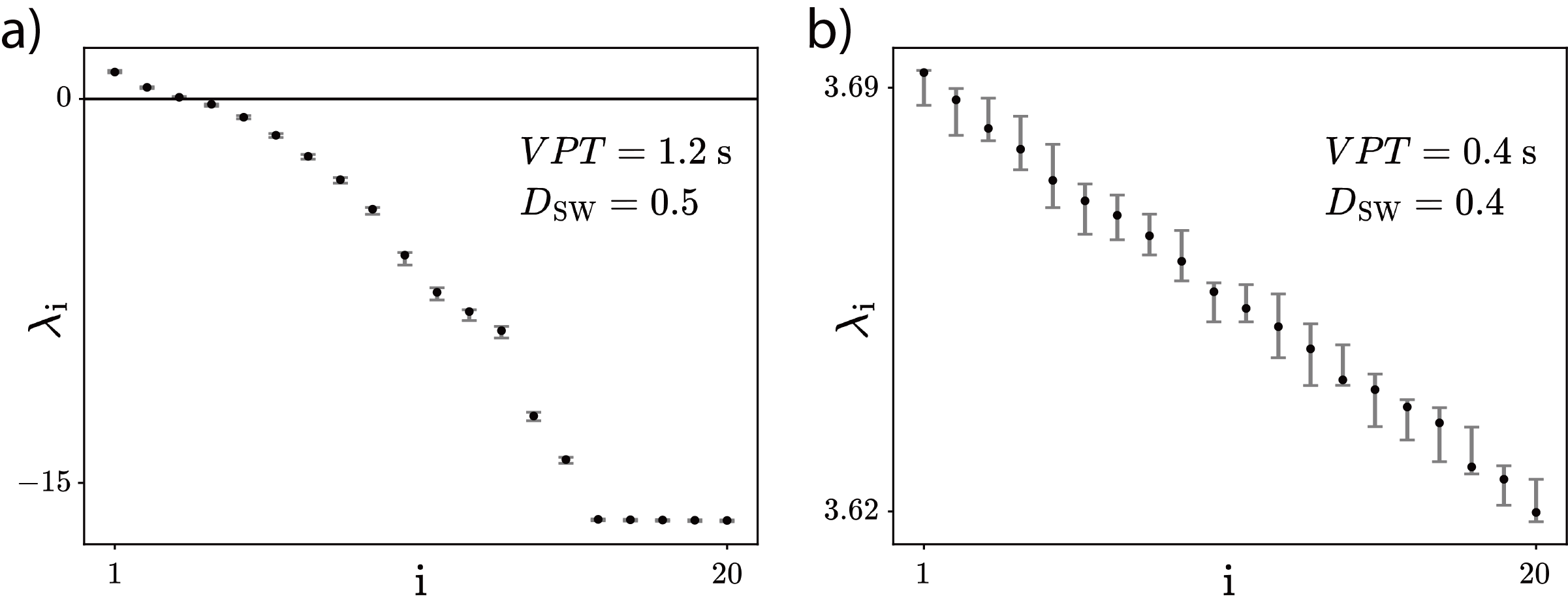}
\caption{{\bf For complex data, the Lyapunov spectrum inferred from our reservoir approach strongly depends on the conditional Lyapunov exponent constraint.} We show the spectrum derived from {\em C. elegans} posture dynamics with (a) and without (b) the constraint. Error bars represent median 95\% confidence intervals bootstrapped from $10^4$ samples from the same individual posture time series as in Fig.~\ref{fig:fig4}.}
\label{fig:S_wrong}
\end{center}
\end{figure}

\begin{table}[h]
  \centering
  \caption{Optimal reservoir hyperparameters across systems}
  \label{tab:hyperparameters}
  \begin{tabular}{lccccc}
  \hline
  \textbf{Parameter} & \textbf{Logistic} & \textbf{Standard} & \textbf{Standard} & \textbf{Double} & \textbf{\textit{C. elegans}} \\
                     & \textbf{map}      & \textbf{(full)}   & \textbf{(partial)} & \textbf{pendulum} & \textbf{L4} \\
  \hline
  \multicolumn{6}{l}{\textit{Network architecture}} \\
  Neurons $N$              &  100   &  100   &  398   & 700 &  10000   \\
  Spectral radius $\rho$   &   0  &   0  &   0.1  & 0.71 &  0.181   \\
  Leaking rate $\alpha$    & 1.0 & 1.0 & 1.0 & 0.17 &  0.706   \\
  Input scaling $\sigma_{\text{in}}$   &  1.0   &  1.0  &  1.0   & 1.0 &  1.267   \\
  Bias scaling $\sigma_{b}$            &  1.0   &  1.0   &  1.0   & 1.0 &  0.01   \\
  Sparsity                 &  0   &   0  &   0  & 0 &   0.99  \\
  Regularization $\alpha_{\text{reg}}$ & 0    &  0   &  0   & 0.01 &  1.5   \\
  \hline
  \multicolumn{6}{l}{\textit{Training}} \\
  Washout steps            &     &     &     & 5000 &  2000   \\
  Training length          &     &     &     &  45000 &  29871   \\
  Sampling interval $\Delta t$ & 1 & 1 & 1 & 0.1 &  1/16   \\
  \hline
  \multicolumn{6}{l}{\textit{Lyapunov spectrum estimation}} \\
  Forcing steps            &     &     &     & 1000 & 500   \\
  Transient steps          &     &     &     & 0 & 500    \\
  Simulation steps            &     &     &     & 5000 & 2500     \\
  Reorthonormalization interval $\tau$ &     &     &  & 1 & 2     \\
  \hline
  \end{tabular}
  \end{table}
  \clearpage

\end{document}